# A consistent *δ-Plus*-ULPH model towards higher accuracy and lower numerical dissipation with fewer neighboring particles

Shi-Xian Wu [a], Peng-Nan Sun [a, b, *], Xiao-Ting Huang [a], Yu-Xiang Peng [a, b], Andrea Colagrossi [c]

[a] *School of Ocean Engineering and Technology, Sun Yat-sen University, Zhuhai, 519000, China*

[b] *Guangdong Provincial Key Laboratory of Information Technology for Deep Water Acoustics, Zhuhai, 519000, China*

[c] *CNR-INM, Institute of Marine Engineering, Rome, 00128, Italy*

* Corresponding author. School of Ocean Engineering and Technology, Sun Yat-sen University, Zhuhai, 519000, China.
  E-mail address: sunpn@mail.sysu.edu.cn (Peng-Nan Sun).

**Highlights**

- A novel consistent $δ^+$-ULPH model is proposed.
- Higher accuracy and lower numerical dissipation with fewer neighbors are achieved.
- Computational efficiency is improved with fewer neighboring particles.
- The proposed model is validated through five benchmarks.

**Abstract**

This paper proposes a novel consistent $δ^+$- Updated Lagrangian Particle Hydrodynamics (ULPH) model. Although the Smoothed Particle Hydrodynamics (SPH) model has gained recognized achievements, it is afflicted by excessive numerical dissipation when the neighboring particles are insufficient. The present proposed consistent $δ^+$-ULPH model has advantages in overcoming this problem. To improve the accuracy, efficiency, stability, and energy conservation, several new techniques are introduced to the consistent $δ^+$-ULPH model. A novel extended support domain technique is proposed to achieve higher accuracy with fewer neighboring particles. An optimal matrix for the velocity divergence is proposed to improve the free-surface stability. A consistent particle shifting technique for the ULPH scheme is proposed to maintain a uniform and regular particle distribution and obtain superior conservation. In addition, an acoustic damper term for the ULPH scheme is introduced to improve the pressure field stability. Five benchmark tests were carried out to validate the consistent $\delta^+$-ULPH model. The conventional ULPH and the consistent $\delta^+$-SPH results are presented for comparison. Results indicate that the proposed consistent $\delta^+$-ULPH model can accurately simulate both gentle waves and violent sloshing flows and shows higher accuracy and lower numerical dissipation when using fewer neighboring particles, even in long-term and long-distance wave propagation simulations. Additionally, the computational efficiency of the consistent $\delta^+$-ULPH model is enhanced visibly because of fewer neighboring particles.

## 1. Introduction

In the field of computational fluid dynamics (CFD), it is acknowledged that the Smoothed Particle Hydrodynamics (SPH) method is one of the most promising mesh-free Lagrangian particle methods [1-3]. Benefiting from its superior advantages in solving large deformation, advection-dominated flow, moving boundary, and multiphase interaction issues, the SPH method develops quickly and has achieved great successes in various areas [4-8], including simulations in both natural phenomena (e.g., tides, floods, and sea waves) and engineering (e.g., paint spraying, liquid pipeline transportation, and jet washing). In particular, the SPH method has been successfully applied to the field of ocean and coastal engineering (e.g., wave generation and propagation [9-11], multiphase flows [12-14], dam-break [15,16], and fluid-structure interactions [17-24]).

Despite the above achievements, the conventional SPH models are afflicted by excessive numerical dissipation when the neighboring particles are insufficient [25,26], especially in long-term (e.g., liquid sloshing) and long-distance wave propagation simulations, which seriously limits the applications of the SPH method. Colagrossi et al. [27] initially reported that increasing the neighboring particle number (i.e., increasing the smoothing length) can improve the energy conservation of the SPH model. Especially in simulations with high Reynolds number, more neighbors are needed to maintain energy conservation. However, in long-distance or large-scale 3D simulations, increasing the neighboring particles causes expensive computational costs, which is also a serious problem. Improving the computational efficiency through High-Performance Computing (HPC) and parallel techniques (e.g., GPU parallel computing acceleration [28-32]) to overcome expensive computational costs is another line. Anyway, this paper concentrates on avoiding the excessive numerical dissipation and improving the computational efficiency with fewer neighboring particles.

To further solve the excessive numerical dissipation, kernel gradient correction (KGC) techniques [26,33-35] were introduced to the SPH governing equations. In addition, based on KGC, Huang et al. [36] proposed a $\delta$-SPH$^C$ model. With a novel format of the discrete pressure gradient, the $\delta$-SPH$^C$ model effectively avoids the excessive numerical dissipation using fewer neighboring particles, even in long-term simulations. Thereafter, Lyu et al. [25] reported the derivation of the $\delta$-SPH$^C$ model and further proposed an improved $\delta$-SPH$^C$ model with higher accuracy. Without any increase in smoothing length, these KGC techniques and $\delta$-SPH$^C$ models gained promising achievements in avoiding numerical dissipation. However, the $\delta$-SPH$^C$ model lacks the particle shifting technique (PST) and tensile instability control (TIC), which significantly limits its applications in violent fluid-structure interactions. For example, the violent vertical sloshing benchmark in Section 4.5 can not be accurately simulated by the $\delta$-SPH$^C$ model. In addition, in $\delta$-SPH$^C$ simulations involving violent fluid impacts and free-surface splashing, numerical instability at the free-surface region may occur.

Recently, based on Peridynamics (PD) theory with a clear physical background in solid mechanics, a new mesh-free Lagrangian particle method named Updated Lagrangian Particle Hydrodynamics (ULPH) was proposed by Tu and Li [37]. Similar to the SPH method, the mesh-free ULPH method has natural advantages in solving large deformation, advection-dominated flow, moving boundary, and multiphase interaction issues. The ULPH method exhibits several additional advantages. Firstly, only the kernel function itself is involved in the discrete ULPH differential operators, while the kernel gradient (which is used in the differential operators of the SPH) is unnecessary. In other words, the requirements for the kernel function selection are reduced. The kernel function of the ULPH model does not need to satisfy the continuity of the first or the second derivatives. Secondly, since the ULPH method is developed from the PD theory, the ULPH model is more easily coupled with solid mechanics and then simulates fluid-structure interactions. Thirdly, an improved conservative pressure gradient term derived from the PD theory is used in the discrete ULPH momentum equation. It will be demonstrated in the present work that, compared to the standard SPH method, the improved pressure gradient term of ULPH contributes to achieving lower discrete error when using a small smoothing length (i.e., fewer neighboring particles). This helps in several aspects, including less noise in the pressure field and less numerical dissipation in the energy view. For example, in the simulation of gravity water wave propagation, the wave attenuation is less. Moreover, fewer neighboring particles mean less computational cost.

The ULPH method has been successfully applied to investigate free-surface flows [38], multiphase flows [39,40], fluid-structure interactions [41,42], and heat conduction [43], demonstrating that the ULPH method shows good accuracy. Despite these achievements, due to the kernel function truncation at the free-surface and its vicinity, free-surface instability in long-term ULPH simulations is still an open problem. This problem will be further reported and addressed in this paper.

Given the aforementioned advantages of the ULPH, our study focuses on developing a consistent $\delta^+$-ULPH model, aiming to achieve higher accuracy and lower numerical dissipation with fewer neighboring particles, particularly in long-term and long-distance wave propagation simulations. In addition, the new model with the employment of PST and TIC also aims to overcome the drawbacks of the $\delta$-SPH$^C$ model (i.e., hard to simulate violent flows with strong negative pressure, which may induce flow voids due to tensile instability). It will be demonstrated that the proposed consistent $\delta^+$-ULPH model can accurately simulate both gentle waves and violent sloshing flows.

This paper is constructed as follows. Section 2 briefly recalls the ULPH scheme. In Section 3, a consistent $\delta^+$-ULPH model is proposed with detailed illustrations. In Section 4, the computational accuracy, efficiency, and conservation of the proposed model are validated through five benchmarks. Finally, Section 5 concludes the present work.

**2. Revisit of the ULPH model**

*2.1. Governing equations of ULPH*

The discrete governing equations of ULPH for free-surface flows are written below [38]:

$$\begin{cases} \dfrac{\mathrm{D}\rho_i}{\mathrm{D}t} = -\rho_i \langle \nabla \cdot \boldsymbol{u} \rangle_i + \Phi_i \\ \dfrac{\mathrm{D}\boldsymbol{u}_i}{\mathrm{D}t} = -\dfrac{1}{\rho_i} \langle \nabla p \rangle_i + \dfrac{1}{\rho_i} \boldsymbol{F}_i^v + \boldsymbol{g} \\ \dfrac{\mathrm{D}\boldsymbol{r}_i}{\mathrm{D}t} = \boldsymbol{u}_i \\ p_i = c_0^2 (\rho_i - \rho_0) \end{cases} \quad (1)$$

where $\rho_i$, $p_i$, $\boldsymbol{u}_i$, and $\boldsymbol{r}_i = (x_i, y_i, z_i)$ denote the density, pressure, velocity vector, and position vector of fluid particle $i$, respectively. $\rho_0$, $c_0$, and $\boldsymbol{g}$ are the initial fluid density, artificial sound speed, and gravity acceleration, respectively. In this paper, the value of $\boldsymbol{g}$ is 9.81 m/s². $\langle \bullet \rangle_i$ represents the particle approximation of $\bullet$.

The discrete differential operators of ULPH are expressed as follows [39]:

$$\begin{cases} \nabla_i (\bullet) = \sum_j W_{ij} \Delta(\bullet) \left( \boldsymbol{M}_i^{-1} \boldsymbol{r}_{ji} \right) V_j \\ \nabla_i \cdot (\bullet) = \sum_j W_{ij} \Delta(\bullet) \cdot \left( \boldsymbol{M}_i^{-1} \boldsymbol{r}_{ji} \right) V_j \end{cases} \quad (2)$$

in which $\Delta(\bullet)$ represents $(\bullet)_j - (\bullet)_i$, where $i$ and $j$ represent the target particle and neighboring particle, respectively. $\boldsymbol{r}_{ji} = \boldsymbol{r}_j - \boldsymbol{r}_i$. $W_{ij}$ is the kernel function. In this paper, the Wendland $C^2$ kernel function with a $2h$ radius is used. $h$ is the smoothing length determined by a multiple $\zeta = h/\Delta x$, where $\Delta x$ is the initial particle distance. A smaller $\zeta$ indicates that fewer neighboring particles are involved in the calculation, resulting in lower computational cost and higher efficiency. The main works in this paper are conducted and discussed with $\zeta = 1.35$ and $2.0$. $V$ is the particle volume. $M_i$ is the moment matrix of $i$, expressed as follows [39]:

$$M_i = \sum_j W_{ij} r_{ji} \otimes r_{ji} V_j \tag{3}$$

Based on Peridynamics theory and the above differential operators, $\langle \nabla \cdot u \rangle_i$ and $\langle \nabla p \rangle_i$ in Eq. (1) can be obtained [38]:

$$\begin{cases} \langle \nabla \cdot u \rangle_i = \sum_j W_{ij} (u_j - u_i) \cdot (M_i^{-1} r_{ji}) V_j \\ \langle \nabla p \rangle_i = \sum_j W_{ij} (p_i M_i^{-1} + p_j M_j^{-1}) r_{ji} V_j \end{cases} \tag{4}$$

In Eq. (1), $F_i^V$ is the viscous term, expressed as follows [38]:

$$F_i^V = \mu K \sum_j W_{ij} \frac{(u_j - u_i) \cdot r_{ji}}{\|r_{ji}\|^2 + (0.1h)^2} (M_i^{-1} r_{ji}) V_j \tag{5}$$

in which $K$ is equal to $2(K_d+2)$, where $K_d$ is the dimension number. The dynamic viscosity $\mu = v \cdot \rho_0$, where $v$ is the kinematic viscosity. It should be noticed that when inertia forces are dominant (i.e., physical viscosity plays a minor role), artificial viscosity can be applied to replace the physical viscosity $F_i^V$ through replacing $\mu K$ by $\alpha h c_0 \rho_0$, where $\alpha$ denotes the coefficient of the artificial viscosity.

To avoid the unphysical pressure oscillations, the density diffusive term $\Phi_i$ is introduced to the continuity equation in Eq. (1), written as follows [38]:

$$\begin{cases} \Phi_i = \delta h c_0 \sum_j W_{ij} \phi_{ij} \cdot (M_i^{-1} r_{ji}) V_j \\ \phi_{ij} = \frac{2(\rho_j - \rho_i) r_{ji}}{\|r_{ji}\|^2} - (\langle \nabla \rho \rangle_i + \langle \nabla \rho \rangle_j) \end{cases} \tag{6}$$

in which $\delta = 0.1$ in this paper.

*2.2. Optimal matrix at the free-surface region*

Considering violent free-surface flows, the kernel function truncation at the free-surface and its vicinity (i.e., the region within a kernel radius away from the free-surface) may cause ill-conditioned of the matrix $M$. According to Yan et al. [38], it is reasonable to assume that the derivative of a specific direction is dominated by this direction, whereas the contribution of other directions is negligible and can be disregarded. Given this assumption, an optimal matrix $M'$ is obtained by neglecting the off-diagonal elements of the original $M$. When calculating the governing equations, the optimal matrix $M'$ is suggested to replace the original $M$ at the free-surface and its vicinity regions. The optimal matrix $M'$ is written below [38]:

$$M'_i = \begin{pmatrix} \sum_j W_{ij}(x_j - x_i)^2 V_j & 0 & 0 \\ 0 & \sum_j W_{ij}(y_j - y_i)^2 V_j & 0 \\ 0 & 0 & \sum_j W_{ij}(z_j - z_i)^2 V_j \end{pmatrix} \tag{7}$$

*2.3. Free-surface detection method*

Free-surface identification is a fundamental procedure for applying the optimal matrix $M'$ and particle shifting technique (PST). Consistent with the identification procedure designed by Marrone et al. [44], the free-surface detection includes two steps as follows:

(i) **Roughly detect the free-surface region.** It exploits the minimum eigenvalue $\lambda$ of the dimensionless detection matrix $M^D$, which is expressed as follows [38]:

$$M_i^D = \sum_j W_{ij} \frac{\mathbf{r}_{ji}}{\Delta x} \otimes \frac{\mathbf{r}_{ji}}{\Delta x} V_j \tag{8}$$

The range of the minimum eigenvalue $\lambda$ is determined by the smoothing length $h$. According to Yan et al. [38], in simulations where $h$ equals $2\Delta x$, the first step of detection can be defined as follows:

$$i \in \begin{cases} F, & \lambda < 0.6 \\ B, & 0.6 \leq \lambda \leq 1 \\ I, & \lambda > 1 \end{cases}, \text{ when } h = 2\Delta x \tag{9}$$

where F and I denote the free-surface and inner region, respectively. B represents the region close to the free-surface or the region with an uneven distribution of particles. When $h$ equals $1.35\Delta x$, the range of $\lambda$ is also tested in this paper. As shown in Fig. 1, a dam-break benchmark is calculated using the ULPH model. We found that $\lambda$ is lower than 0.3 and tends to 0 for free-surface particles, whereas $\lambda$ tends to 0.5 for inner particles. This range of $\lambda$ is consistent in both 2D and 3D simulations. Therefore, Eq. (9) should be altered as follows:

$$i \in \begin{cases} \text{F}, & \lambda < 0.3 \\ \text{B}, & 0.3 \leq \lambda \leq 0.45 \\ \text{I}, & \lambda > 0.45 \end{cases}, \quad \text{when } h = 1.35\Delta x \tag{10}$$

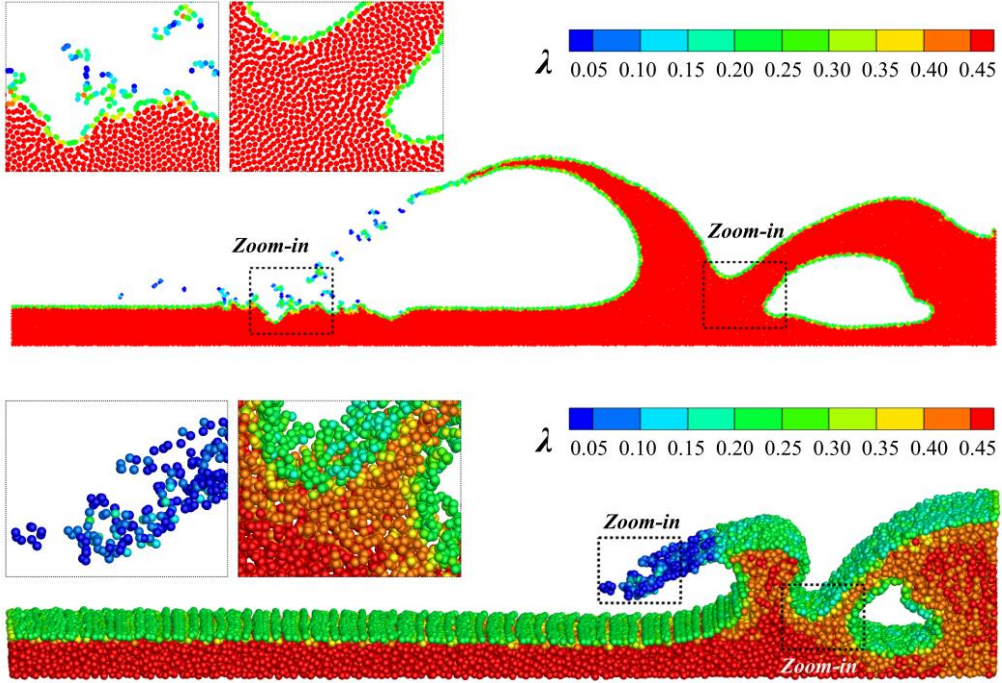

Fig. 1. The minimum eigenvalue $\lambda$ calculated by the ULPH model ($h = 1.35\Delta x$). Top: 2D dam-break; Bottom: 3D dam-break.

(ii) **Further detect by searching an "umbrella-shaped" region**. Considering the particles that belong to $\text{B}$, a further identification is needed. The unit normal vector is evaluated first, written as [38]:

$$\boldsymbol{n}_i = \frac{\langle \nabla \lambda \rangle_i}{\|\langle \nabla \lambda \rangle_i\|}, \quad \langle \nabla \lambda \rangle_i = -\sum_j W_{ij}\left(\lambda_j - \lambda_i\right) \boldsymbol{M}_i^{-1} \boldsymbol{r}_{ji} V_j \tag{11}$$

Subsequently, according to the normal vector direction, the "umbrella-shaped" region is determined. If there are no neighboring particles in the "umbrella-shaped" region, the concerned particle is at the free-surface. More details about the "umbrella-shaped" region can be found in the reference [44].

## 3. Consistent $\delta^+$-ULPH model

### 3.1. Governing equations of consistent $\delta^+$-ULPH

Encouraged by the consistent $\delta^+$-SPH model [45], adopting the particle shifting technique (PST) [45], tensile instability control (TIC) technique [41,46], and acoustic damper term [47], a weakly-compressible consistent $\delta^+$-ULPH model is constructed as follows:

$$\begin{cases} \dfrac{d\rho_i}{dt} = -\rho_i \langle \nabla \cdot \boldsymbol{u} \rangle_i - \rho_i \langle \nabla \cdot \delta\boldsymbol{u} \rangle_i + \langle \nabla \cdot (\rho\delta\boldsymbol{u}) \rangle_i + \Phi_i \\ \dfrac{d\boldsymbol{u}_i}{dt} = -\dfrac{1}{\rho_i} \langle \nabla p \rangle_i + \dfrac{1}{\rho_i} \langle \nabla \cdot (\rho\boldsymbol{u} \otimes \delta\boldsymbol{u}) \rangle_i + \dfrac{1}{\rho_i} \boldsymbol{F}_i^v + \dfrac{1}{\rho_i} \boldsymbol{F}_i^{ad} + \boldsymbol{g} \\ \dfrac{d\boldsymbol{r}_i}{dt} = \boldsymbol{u}_i + \delta\boldsymbol{u}_i \\ p_i = c_0^2(\rho_i - \rho_0) \end{cases} \quad (12)$$

where $\delta\boldsymbol{u}_i$ is the shifting velocity. $\langle \nabla \cdot \delta\boldsymbol{u} \rangle_i$, $\langle \nabla \cdot (\rho\delta\boldsymbol{u}) \rangle_i$, and $\langle \nabla \cdot (\rho\boldsymbol{u} \otimes \delta\boldsymbol{u}) \rangle_i$ are the $\delta\boldsymbol{u}$-terms, which will be further discussed in Section 3.4. $\boldsymbol{F}_i^{ad}$ is the acoustic damper term, which will be further reported in Section 3.6.

In the discretization of Eq. (12), matrix $\boldsymbol{M}$ in Eq. (3) is used. However, $\boldsymbol{M}$ may become ill-conditioned for splashing and isolated particles, as a result of the serious lack of neighboring particles. To avoid this, when the minimum eigenvalue $\lambda$ of $\boldsymbol{M}_i^D$ in Eq. (8) is $< 0.2$ (where the corresponding smoothing length is $h = 2\Delta x$. For different smoothing lengths and kernel functions, the value 0.2 can be altered), the discrete governing equations of particle $i$ and its neighboring particles are switched to the SPH scheme where the matrix $\boldsymbol{M}$ does not need, i.e., calculating the velocity divergence and pressure gradient using the kernel gradient operator. This is effective since the splashing and isolated particles can be effectively identified by $\lambda$. It should be noticed that in multi-phase flows (i.e., considering the liquid and air phases), the support domain is always complete and the matrix $\boldsymbol{M}$ is well-conditioned, meaning that this switch is not required.

*3.2. Extended support domain technique*

As highlighted in Section 1, compared to the standard SPH method, an improved pressure gradient term of ULPH helps to achieve lower discrete error when using a small smoothing length. We present an illustration in this subsection.

According to Monaghan [48,49], the discrete pressure gradient of the standard SPH model can be expressed as follows:

$$\langle \nabla p \rangle_i^{SPH} = \rho_i \sum_j \left( \frac{p_i}{\rho_i^2} + \frac{p_j}{\rho_j^2} \right) \nabla W_{ij} m_j \qquad (13)$$

To accurately and stably simulate multi-phase flows with large density differences, another discrete approximation of the pressure gradient is used [50]:

$$\langle \nabla p \rangle_i^{SPH} = \sum_j (p_i + p_j) \nabla W_{ij} V_j \qquad (14)$$

Derived from the PD theory (see Appendix A), the discrete pressure gradient of the ULPH model introduces a tensor correction, where a form of $(p_i \boldsymbol{M}_i^{-1} + p_j \boldsymbol{M}_j^{-1})$ is used, instead of $(p_i + p_j)$ in standard SPH, written as:

$$\begin{cases} \langle \nabla p \rangle_i^{ULPH} = \sum_j W_{ij} \left( p_i \boldsymbol{M}_i^{-1} + p_j \boldsymbol{M}_j^{-1} \right) \boldsymbol{r}_{ji} V_j \\ \boldsymbol{M}_i = \sum_j W_{ij} \boldsymbol{r}_{ji} \otimes \boldsymbol{r}_{ji} V_j \\ \boldsymbol{M}_j = \sum_k W_{jk} \boldsymbol{r}_{kj} \otimes \boldsymbol{r}_{kj} V_k \end{cases} \qquad (15)$$

in which particles *i*, *j*, and *k* are used, while only particles *i* and *j* are involved in the standard SPH in Eqs. (13) and (14). Particles indexed by *k* are the neighbours of particle *j*, which means more particles can support the ULPH calculation. In other words, the support domain seems to be extended without changing the smoothing length. Therefore, we named this "extended support domain technique (ESDT)". This allows to achieve higher accuracy at the same smoothing length, because more supporting particles mean less discrete error in the ULPH. From another perspective, similar accuracy to the SPH can be achieved by the ULPH with a smaller smoothing length (i.e., fewer neighboring particles). Meanwhile, fewer neighboring particles mean lower computational costs and higher efficiency.

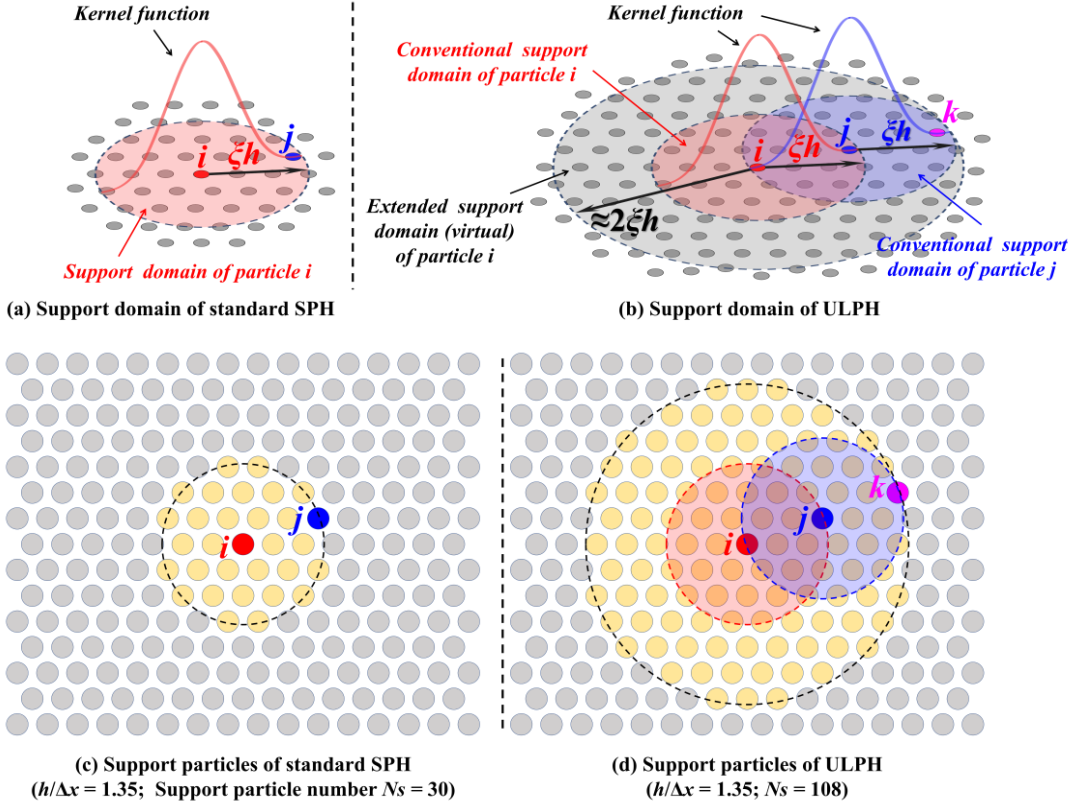

Fig. 2. Illustration of the extended support domain by comparing the standard SPH and ULPH.

As shown in Fig. 2, we further illustrate the ESDT by comparing the support domain and support particle number between the SPH and ULPH. $i$ represents the target particle, $j$ is one of the outermost neighboring particles of $i$, and $k$ is one of the neighboring particles of $j$. Considering a single step calculation, only the neighboring particles of $i$ (see the red region in Fig. 2 (a)) are used to calculate the SPH pressure gradient in Eqs. (13) and (14), despite the pressure $p_j$ includes the effect of its surrounding particles in former steps. While for the ULPH pressure gradient in Eq. (15), since the matrix $M_j^{-1}$ is calculated by the neighboring particles of $j$ (see the blue region in Fig. 2 (b)) in each step, more particles (see the gray region in Fig. 2 (b)) support the calculation without changing the smoothing length. This indicates that the support domain radius of $i$ seems to be extended from the original $\xi h$ to approximately $2\xi h$. In this paper, $\xi = 2$ is used with the Wendland $C^2$ kernel function. As shown in Fig. 2 (c) and (d), $Ns$ represents the number of supporting particles, which participate and support the calculation, including but not only the neighboring particles in ULPH. With $h/\Delta x = 1.35$, $Ns = 30$ in the standard SPH model, while $Ns = 108$ in the ULPH model.

To improve the accuracy and conservation, in the proposed consistent $\delta^+$-ULPH model, a symmetry matrix $(M_i^{-1} + M_j^{-1})/2$ is used in the numerical diffusive terms, instead of a single matrix $M_i^{-1}$ in Eqs. (5) and (6). This is another implementation of ESDT, since the matrix $M_j^{-1}$ enables more particles to support the discrete approximations without changing the smoothing length. Using this symmetry matrix, the density diffusive term $\Phi_i$ in Eq. (12) is expressed as follows:

$$\begin{cases} \Phi_i = \delta h c_0 \sum_j W_{ij} \phi_{ij} \cdot \left( \dfrac{M_i^{-1} + M_j^{-1}}{2} r_{ji} \right) V_j \\ \phi_{ij} = \dfrac{2(\rho_j - \rho_i) r_{ji}}{\|r_{ji}\|^2} - \left( \langle \nabla \rho \rangle_i + \langle \nabla \rho \rangle_j \right) \\ \langle \nabla \rho \rangle_i = \sum_j W_{ij} (\rho_j - \rho_i) \dfrac{M_i^{-1} + M_j^{-1}}{2} r_{ji} V_j \end{cases} \quad (16)$$

Similarly, the viscous term $F_i^v$ in Eq. (12) is expressed as follows:

$$F_i^v = \mu K \sum_j W_{ij} \frac{(u_j - u_i) \cdot r_{ji}}{\|r_{ji}\|^2 + (0.1h)^2} \left( \frac{M_i^{-1} + M_j^{-1}}{2} r_{ji} \right) V_j \quad (17)$$

The use of the symmetry matrix $(M_i^{-1} + M_j^{-1})/2$ also takes into consideration the conservation property. When the continuity equation in Eq. (12) integrates, the integration of $\Phi_i$ in Eq. (16) equals 0 thanks to the anti-symmetry property, which ensures mass conservation. Similarly, the integration of $F_i^v$ in Eq. (17) equals 0 and therefore ensures the momentum conservation. A detailed explanation about the conservation is presented in Appendix B.

In summary, the ESDT is a crucial technique for the consistent $\delta^+$-ULPH model to achieve superior accuracy, efficiency, and conservation with fewer neighboring particles.

*3.3. Optimal matrix for the velocity divergence*

According to Yan et al. [38], the calculation accuracy is significantly influenced by the matrix $M$. However, in single-phase ULPH simulations, the matrix $M$ at the free-surface and its vicinity may become ill-conditioned, caused by the kernel function truncation. This further results in free-surface instability in long-term simulations. In this subsection, an optimal matrix $\boldsymbol{D}$ for the velocity divergence is defined to improve the free-surface stability. The derivation of matrix $\boldsymbol{D}$ is as follows.

In Eq. (12), the conventional ULPH discrete scheme of $\langle \nabla \cdot \boldsymbol{u} \rangle_i$ is written as:

$$\langle \nabla \cdot \boldsymbol{u} \rangle_i^{ULPH} = \sum_j (\boldsymbol{u}_j - \boldsymbol{u}_i) \cdot \left[ W_{ij} \boldsymbol{M}_i^{-1} (\boldsymbol{r}_j - \boldsymbol{r}_i) \right] V_j \tag{18}$$

in which,

$$W_{ij} \boldsymbol{M}_i^{-1} (\boldsymbol{r}_j - \boldsymbol{r}_i) = W_{ij} \begin{pmatrix} M_{i11}^{-1}(x_j - x_i) + M_{i12}^{-1}(y_j - y_i) + M_{i13}^{-1}(z_j - z_i) \\ M_{i21}^{-1}(x_j - x_i) + M_{i22}^{-1}(y_j - y_i) + M_{i23}^{-1}(z_j - z_i) \\ M_{i31}^{-1}(x_j - x_i) + M_{i32}^{-1}(y_j - y_i) + M_{i33}^{-1}(z_j - z_i) \end{pmatrix} \tag{19}$$

Referring to the SPH scheme, the discrete form of $\langle \nabla \cdot \boldsymbol{u} \rangle_i$ is defined as:

$$\langle \nabla \cdot \boldsymbol{u} \rangle_i^{SPH} = \sum_j (\boldsymbol{u}_j - \boldsymbol{u}_i) \cdot \nabla_i W_{ij} V_j \tag{20}$$

where,

$$\nabla_i W_{ij} = \frac{dW_{ij}}{dr_{ij}} \frac{\boldsymbol{r}_j - \boldsymbol{r}_i}{r_{ij}} = \frac{dW_{ij}}{r_{ij} dr_{ij}} \begin{pmatrix} x_j - x_i & y_j - y_i & z_j - z_i \end{pmatrix}^T \tag{21}$$

in which $r_{ij} = \|\boldsymbol{r}_j - \boldsymbol{r}_i\|$. Since the SPH discrete scheme shows good stability at the free-surface region, it is reasonable to construct a similar form for the ULPH scheme. It can be noticed that the $W_{ij} \boldsymbol{M}_i^{-1} (\boldsymbol{r}_j - \boldsymbol{r}_i)$ in Eq. (18) and the kernel gradient $\nabla_i W_{ij}$ in Eq. (20) exert similar effects. Given this, it's reasonable to modify the $W_{ij} \boldsymbol{M}_i^{-1} (\boldsymbol{r}_j - \boldsymbol{r}_i)$ to align with the form of $\nabla_i W_{ij}$. Through neglecting the off-diagonal elements and taking the average of the main diagonal elements of $\boldsymbol{M}_i$, the diagonal optimal matrix $\boldsymbol{D}_i$ of the velocity divergence is obtained:

$$\begin{cases} \boldsymbol{D}_i = \begin{pmatrix} \mathscr{D} & 0 & 0 \\ 0 & \mathscr{D} & 0 \\ 0 & 0 & \mathscr{D} \end{pmatrix} \\ \mathscr{D} = \frac{1}{3} \sum_j W_{ij} \left[ (x_j - x_i)^2 + (y_j - y_i)^2 + (z_j - z_i)^2 \right] V_j \end{cases} \tag{22}$$

As a result, Eq. (19) is rewritten as:

$$W_{ij} \boldsymbol{D}_i^{-1} (\boldsymbol{r}_j - \boldsymbol{r}_i) = W_{ij} \frac{1}{\mathscr{D}} \begin{pmatrix} x_j - x_i & y_j - y_i & z_j - z_i \end{pmatrix}^T \tag{23}$$

Eventually, $\langle \nabla \cdot \boldsymbol{u} \rangle_i$ in Eq. (12) is expressed as:

$$\langle \nabla \cdot \boldsymbol{u} \rangle_i = \sum_j W_{ij} (\boldsymbol{u}_j - \boldsymbol{u}_i) \cdot (\boldsymbol{D}_i^{-1} \boldsymbol{r}_{ji}) V_j \tag{24}$$

It should be noticed that the diagonal matrix $\boldsymbol{D}_i$ is applied to the divergence calculation of the physical velocity and shifting velocity for the whole fluid domain, while other discrete differential terms in Eq. (12) still employ the matrix $\boldsymbol{M}$. In addition, the optimal matrix $\boldsymbol{M}'$ in Eq. (7) is also applied for the free-surface region in Eq. (12) (except for the discrete terms applying matrix $\boldsymbol{D}$), as illustrated in Section 2.2. Applying the matrix $\boldsymbol{M}'$ to the consistent $\delta^+$-ULPH model also needs free-surface detection. The detection procedure has been illustrated in Section 2.3.

### 3.4. Particle shifting technique

In the particle-based methods, the particle shifting technique (PST) is able to address the tensile instability and particle aggregation problems and obtain a uniform and regular particle distribution. Xu et al. [51] originally proposed the PST, which was developed by Lind et al. [52] and Sun et al. [45,53]. Recently, an improved PST (IPST) has been proposed by Wang et al. [54]. Even in long-term simulations, the free-surface and its vicinity maintain a uniform particle distribution after applying the IPST.

Referring to the IPST [54], the shifting velocity $\delta \hat{\boldsymbol{u}}_i$ of the consistent $\delta^+$-ULPH model is expressed as follows:

$$\delta \hat{\boldsymbol{u}}_i = \begin{cases} -\text{Ma} \cdot 2h \cdot c_0 \sum_j W_{ij} (1 + \chi_{ij}) (\boldsymbol{M}_i^{-1} \boldsymbol{r}_{ji}) V_j, & \text{if } i \in \text{I}_1 \\ -\text{Ma} \cdot 2h \cdot c_0 \sum_j W_{ij} (1 + \chi_{ij}) \left( \dfrac{\boldsymbol{M}_i^{-1} + \boldsymbol{M}_j^{-1}}{2} \boldsymbol{r}_{ji} \right) V_j, & \text{if } i \in \text{I}_2 \\ 0, & \text{if } i \in \text{F} \\ \delta \boldsymbol{u}_i^*, & \text{if } i \in \text{V} \end{cases} \tag{25}$$

in which $\chi_{ij} = 0.2 (W_{ij} / W(\Delta x))^{0.4}$. Ma = $U_{\max}/c_0$, where $U_{\max}$ is the maximum expected fluid velocity. F and V denote the free-surface and its vicinity regions, respectively. $\text{I}_1$ and $\text{I}_2$ denote the inner region, as illustrated in Fig. 3.

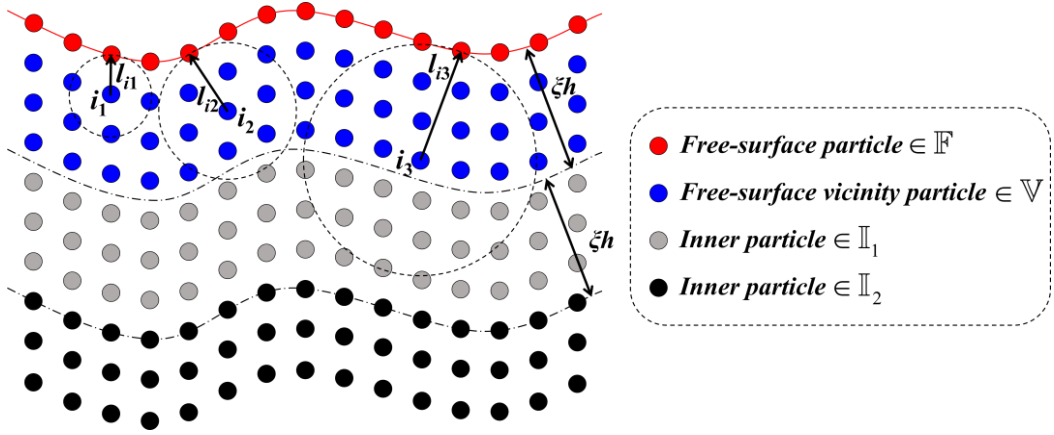

Fig. 3. Sketch of the free-surface, free-surface vicinity, and inner particles. The concerned domain of the calculation of the shifting vectors within the free-surface vicinity is illustrated.

According to Colagrossi et al. [55], the unevenness in the particle distribution is measured by the vector $\nabla \Gamma_i$, and the particle-lacking regions are always pointed towards by the vector $-\nabla \Gamma_i$. Referring to the SPH scheme, we found that $\boldsymbol{N}_i$ is effective to point towards the particle-lacking regions in the ULPH framework. However, at the free-surface region, $-\nabla \Gamma_i$ and $\boldsymbol{N}_i$ always point to the outside of the fluid and the shifting direction directly determined by $\boldsymbol{N}_i$ is improper. Therefore, to maintain a reasonable shifting direction, a variable domain of concern is used to calculate $\delta \boldsymbol{u}_i^*$ in Eq. (25). The $\nabla \Gamma_i$, $\boldsymbol{N}_i$, and $\delta \boldsymbol{u}_i^*$ are expressed as follows:

$$\begin{cases} -\nabla \Gamma_i = -\sum_j \nabla_i W_{ij} V_j \\ \boldsymbol{N}_i = -\sum_j W_{ij} \boldsymbol{M}_i^{-1} \boldsymbol{r}_{ji} V_j \end{cases} \quad (26)$$

and

$$\delta \boldsymbol{u}_i^* = -\mathrm{Ma} \cdot 2h \cdot c_0 \sum_j W_{ij} \chi_{ij} \left( \boldsymbol{M}_i^{-1} \boldsymbol{r}_{ji} \right) V_j, \ j \in \left[ \left| \boldsymbol{r}_{ji} \right| < l_i \right] \quad (27)$$

in which $l_i$ is the smallest distance between $i$ and the free-surface particles (see Fig. 3). Before calculating $\delta \boldsymbol{u}_i^*$, an extra loop is necessary for the calculation of $l_i$. The extra computational cost is directly related to the free-surface proportion. The symmetry matrix $\left( \boldsymbol{M}_i^{-1} + \boldsymbol{M}_j^{-1} \right)/2$ is not used to calculate the $\delta \hat{\boldsymbol{u}}_i$ when the target particle $i$ belongs to $\mathbb{I}_1$ and $\mathbb{V}$, given the kernel function truncation in these regions may lower the precision of the matrix $\boldsymbol{M}_j^{-1}$, thereby amplifying the errors.

For the consideration of robustness, a restriction on the shifting value is applied. Eventually, the $\delta \boldsymbol{u}_i$ in Eq. (12) is written as follows:

$$\delta \boldsymbol{u}_i = \min\left(\|\delta \hat{\boldsymbol{u}}_i\|, \frac{U_{\max}}{2}\right) \frac{\delta \hat{\boldsymbol{u}}_i}{\|\delta \hat{\boldsymbol{u}}_i\|} \tag{28}$$

The discrete $\delta \boldsymbol{u}$-terms in Eq. (12) are written below:

$$\begin{cases} \langle \nabla \cdot \delta \boldsymbol{u} \rangle_i = \sum_j W_{ij} \left(\delta \boldsymbol{u}_j - \delta \boldsymbol{u}_i\right) \cdot \left(\boldsymbol{D}_i^{-1} \boldsymbol{r}_{ji}\right) V_j \\ \langle \nabla \cdot (\rho \delta \boldsymbol{u}) \rangle_i = \sum_j W_{ij} \left(\rho_j \delta \boldsymbol{u}_j + \rho_i \delta \boldsymbol{u}_i\right) \cdot \left(\frac{\boldsymbol{M}_i^{-1} + \boldsymbol{M}_j^{-1}}{2} \boldsymbol{r}_{ji}\right) V_j \\ \langle \nabla \cdot (\rho \boldsymbol{u} \otimes \delta \boldsymbol{u}) \rangle_i = \sum_j W_{ij} \left(\rho_j \boldsymbol{u}_j \otimes \delta \boldsymbol{u}_j + \rho_i \boldsymbol{u}_i \otimes \delta \boldsymbol{u}_i\right) \cdot \left(\frac{\boldsymbol{M}_i^{-1} + \boldsymbol{M}_j^{-1}}{2} \boldsymbol{r}_{ji}\right) V_j \end{cases} \tag{29}$$

where the $\langle \nabla \cdot \delta \boldsymbol{u} \rangle_i$ is expressed through a discrete difference form with the optimal matrix $\boldsymbol{D}$ in Eq. (22), given the consistency with $\langle \nabla \cdot \boldsymbol{u} \rangle_i$ in Eq. (24). Conversely, the discrete form of the other two $\delta \boldsymbol{u}$-terms is defined through a sum, with using the symmetry matrix $\left(\boldsymbol{M}_i^{-1} + \boldsymbol{M}_j^{-1}\right)/2$. The latter aims to construct an anti-symmetry form to ensure the conversation property. In light of the anti-symmetry property, the integrations of $\langle \nabla \cdot (\rho \delta \boldsymbol{u}) \rangle_i$ and $\langle \nabla \cdot (\rho \boldsymbol{u} \otimes \delta \boldsymbol{u}) \rangle_i$ are both equal to 0. A detailed explanation about the conservation is presented in Appendix B.

*3.5. Tensile instability control*

In simulations characterized by a large area of negative pressure, tensile instability and numerical cavitation may occur, which seriously limit the applications of conventional SPH and ULPH models. Proposed by Sun et al. [46], the tensile instability control (TIC) technique effectively addresses this problem through a simple switch of the discrete form of the pressure gradient. Recently, Yan et al. [41] have introduced TIC into the ULPH model to investigate water entry problems. After applying the TIC technique, the $\langle \nabla p \rangle_i$ in Eq. (12) is expressed as follows [41]:

$$\langle \nabla p \rangle_i^{TIC} = \begin{cases} \sum_j W_{ij} \left(p_j \boldsymbol{M}_j^{-1} + p_i \boldsymbol{M}_i^{-1}\right) \boldsymbol{r}_{ji} V_j, & p_i \geq 0 \text{ or } i \in (\text{F} \cup \text{V}) \\ \sum_j W_{ij} \left(p_j \boldsymbol{M}_j^{-1} - p_i \boldsymbol{M}_i^{-1}\right) \boldsymbol{r}_{ji} V_j, & \text{else} \end{cases} \tag{30}$$

It is worth noticing that the integration of TIC induces small errors for the momentum conservation (the error can be much reduced by using PST [46]). As a result, to rigorously analyze the conservation property of the numerical models, except for the vertical sloshing benchmark test with strong negative pressures in Section 4.5, TIC is not applied in other benchmark tests.

*3.6. Acoustic damper term*

The weakly-compressible fluid assumption and liquid impacts cause the acoustic component of the pressure field, which weakens the pressure field stability of the weakly-compressible SPH models. To overcome this problem, Sun et al. [47] introduced an acoustic damper term into the SPH model. Referring to the SPH scheme, the acoustics damper term of the weakly-compressible ULPH model in Eq. (12) is expressed as follows:

$$\begin{cases} \boldsymbol{F}_i^{ad} = \alpha_2 h c_0 \rho_0 \langle \nabla \mathcal{D} \rangle_i = \alpha_2 h c_0 \rho_0 \sum_j W_{ij} \left( \mathcal{D}_j + \mathcal{D}_i \right) \left( \frac{\boldsymbol{M}_i^{-1} + \boldsymbol{M}_j^{-1}}{2} \boldsymbol{r}_{ji} \right) V_j \\ \mathcal{D}_i = \langle \nabla \cdot \boldsymbol{u} \rangle_i = \sum_j W_{ij} \left( \boldsymbol{u}_j - \boldsymbol{u}_i \right) \cdot \left( \boldsymbol{D}_i^{-1} \boldsymbol{r}_{ji} \right) V_j \end{cases} \quad (31)$$

where the discrete form of $\langle \nabla \cdot \boldsymbol{u} \rangle_i$ is consistent with Eq. (24). It's worth noting that in simulations involving violent fluid splashing characterized by large deformation and breaking of the free-surface (e.g., violent vertical sloshing benchmark in Section 4.5), the $\mathcal{D}_i$ at free-surface and its vicinity is set to 0, given that the kernel function truncation at the above regions may result in low accuracy of $\mathcal{D}_i$ and then cause the free-surface instability. The dimensionless coefficient $\alpha_2$ of the acoustics damper term is set to 1.0 in this paper as suggested by Sun et al. [47].

*3.7. Time integration*

To ensure computational stability, the time step $\Delta t$ is determined by [47]:

$$\begin{cases} \Delta t_c = K_c \frac{h}{c_0}, \quad \Delta t_v = 0.125 \frac{h^2}{v}, \quad \Delta t_a = 0.25 \min_i \sqrt{\frac{h}{\|\boldsymbol{a}_i\|}}, \quad \Delta t_{ad} = \frac{K_c}{\alpha_2} \frac{h}{c_0} \\ \Delta t = \min \left( \Delta t_c, \Delta t_v, \Delta t_a, \Delta t_{ad} \right) \end{cases} \quad (32)$$

in which $K_c$ is the Courant-Friedrichs-Lewy number. $K_c = 1.2$ is adopted in this paper unless otherwise specified. The fourth-order Runge–Kutta (RK4) scheme is used to integrate the system.

## 4. Numerical results

In this section, computational accuracy, efficiency, and energy conservation of the consistent $\delta^+$-ULPH model with fewer neighboring particles are validated through five benchmarks. To show the superiority of the consistent $\delta^+$-ULPH model, the results of the conventional ULPH model [38] and the consistent $\delta^+$-SPH model [45] are computed for comparison.

### 4.1. Benchmark test No.1: Rotation of a square fluid

Characterized by large free-surface deformation, the rotation of a free-surface square fluid is a challenging test. It is widely used to verify the stability of the meshfree method. A two-dimensional square fluid is set up with a side length $L = 1$ m. The initial velocity field is defined as:

$$\boldsymbol{u}_0(x, y) = (\omega y \quad -\omega x)^T \tag{33}$$

where $\omega$ is the angular velocity and equals 20 rad/s in this paper. Assuming that the fluid is incompressible, the initial pressure field is determined by solving the associated Poisson equation, written as [56]:

$$p_0(x,y) = \rho_0 \sum_{m}^{\infty} \sum_{n}^{\infty} \frac{-32\omega^2}{mn\pi^2 \left[(m\pi/L)^2 + (n\pi/L)^2\right]} \sin\left(\frac{m\pi(x+L/2)}{L}\right) \sin\left(\frac{n\pi(y+L/2)}{L}\right), \quad m, n \in \mathrm{N}_{odd}$$

(34)

in which $\rho_0 = 1000$ kg/m$^3$. Since the series provided by Eq. (34) converges rapidly, the values of $p_0$ may be completely approximated by $m = 1, 3, 5$ and $n = 1, 3, 5$. In this benchmark, $c_0 = 10\sqrt{2}\omega L$ and $\alpha = 0.05$.

To ensure the accuracy of the tests, the convergence of the consistent $\delta^+$-ULPH model ($h/\Delta x = 1.35$) is studied through the domain center pressure using three different particle resolutions ($L/\Delta x = 50, 100,$ and $200$), as shown in Fig. 4. The BEM-MEL result in the reference [56] is also plotted for comparison. It can be observed that the time histories of the center pressure become more consistent and approach the BEM-MEL result as the particle resolution increases. When $t\omega > 3.0$, the results of $L/\Delta x = 100$ and $200$ are nearly identical, even though there is a slight difference when $t\omega < 1.0$. This indicates that the numerical results have converged. In the following tests of this benchmark, the particle resolution $L/\Delta x = 200$ is adopted.

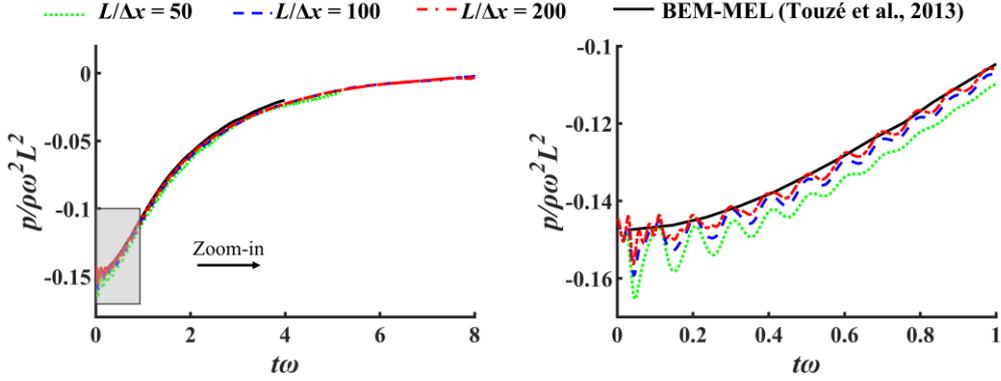

Fig. 4. Rotation of a square fluid: The center pressure evolutions using the consistent $\delta^+$-ULPH model ($h/\Delta x = 1.35$) with three different resolutions ($L/\Delta x = 50, 100,$ and $200$), compared with the BEM-MEL result [56].

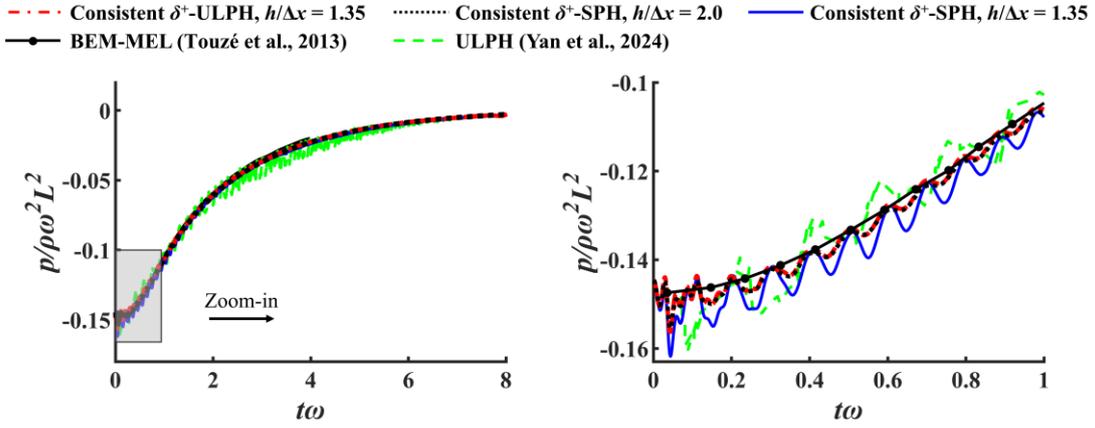

Fig. 5. Rotation of a square fluid: The center pressure evolutions using the consistent $\delta^+$-ULPH model ($h/\Delta x = 1.35$, $L/\Delta x = 200$) and the consistent $\delta^+$-SPH model with acoustics damper ($h/\Delta x = 1.35$ and $2.0$, $L/\Delta x = 200$). The results of the BEM-MEL [56] and the conventional ULPH model [38] are also plotted.

Figure 5 shows the center pressure evolutions calculated by the consistent $\delta^+$-ULPH model and the consistent $\delta^+$-SPH model with the acoustic damper term. For comparison, the conventional ULPH result in reference [38] is also plotted. Observing Fig. 5, when $0 < t\omega < 1.0$, the pressures of particle-based models have slightly unphysical oscillations due to the transition of particle distribution from lattice to uniform. When $2.0 < t\omega < 6.0$, the pressure of the conventional ULPH model shows large high-frequency oscillations. Conversely, the consistent $\delta^+$-ULPH model ($h/\Delta x = 1.35$) and the consistent $\delta^+$-SPH model ($h/\Delta x = 2.0$) have almost no pressure oscillations attributed to the acoustics damper. The consistent $\delta^+$-ULPH model ($h/\Delta x = 1.35$) and the consistent $\delta^+$-SPH model ($h/\Delta x = 2.0$) show equivalent accuracy. However, the accuracy of the latter decreases, and pressure oscillations amplify as the value of $h/\Delta x$ reduces to 1.35.

Figure 6 presents the free-surface evolutions and pressure field calculated by the consistent $\delta^+$-ULPH model, compared with the FDM and BEM-MEL results [56]. The free-surface profile obtained by the consistent $\delta^+$-ULPH model shows good agreement with the FDM and BEM-MEL results, and the pressure field is free of noise. Figure 7 further compares the pressure fields of the consistent $\delta^+$-ULPH and the consistent $\delta^+$-SPH models. The particle resolutions $L/\Delta x$ are consistent. The result shows that the consistent $\delta^+$-SPH model ($h/\Delta x$ = 1.35) exhibits visible pressure perturbations, which reduce as the smoothing length increases. Conversely, the pressure field of the consistent $\delta^+$-ULPH model is fairly good.

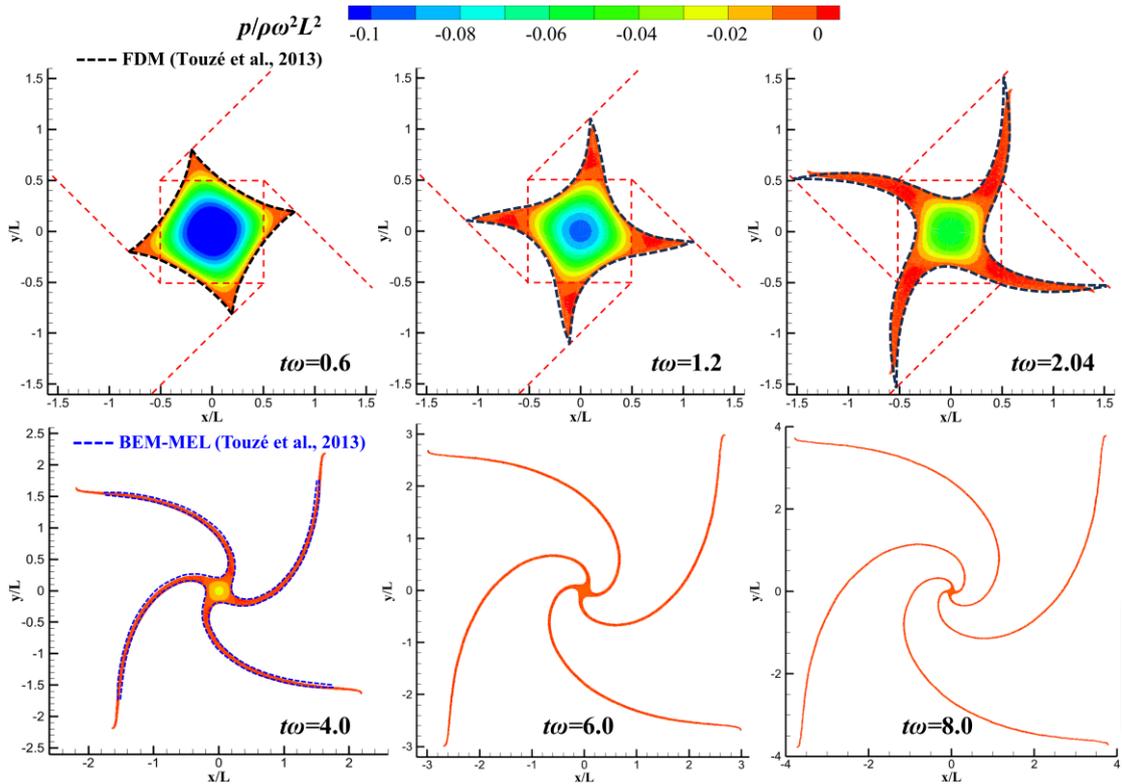

Fig. 6. Rotation of a square fluid: Pressure field evolution using the consistent $\delta^+$-ULPH model ($h/\Delta x$ = 1.35, $L/\Delta x$ = 200). The black and blue dashed lines are the free-surface profiles calculated by the FDM solver [56] and BEM-MEL solver [56], respectively.

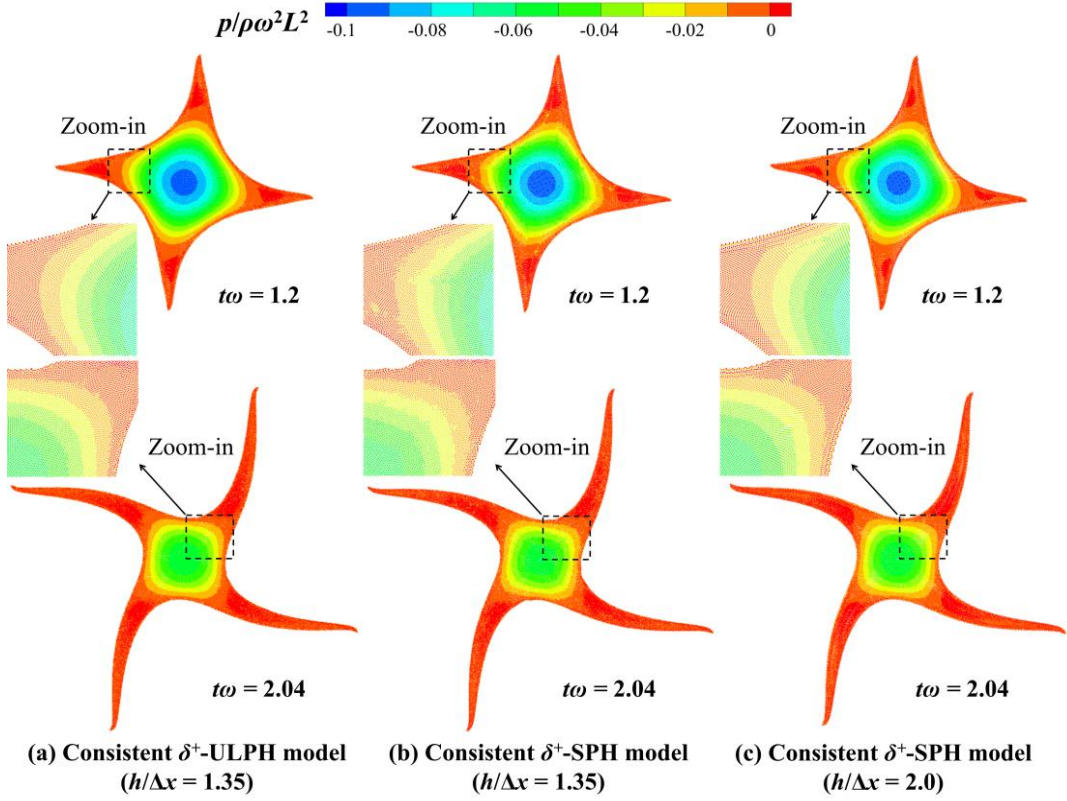

Fig. 7. Rotation of a square fluid: Pressure field comparison between the consistent $\delta^+$-ULPH and the consistent $\delta^+$-SPH results ($L/\Delta x = 200$).

Table 1. Rotation of a square fluid: Computational efficiency comparison of the consistent $\delta^+$-ULPH and consistent $\delta^+$-SPH models. The computational cost is considered for the first 1,000 steps.

| Particle resolution | Particle number | The consistent $\delta^+$-SPH (left: $h/\Delta x = 1.35$; right: $h/\Delta x = 2$) | | | | The consistent $\delta^+$-ULPH ($h/\Delta x = 1.35$) | | |
|---|---|---|---|---|---|---|---|---|
| | | Computational cost (s) | FPS | Computational cost (s) | FPS | Computational cost (s) | FPS | Speed-up rate (%) |
| $L/\Delta x = 50$ | 2,601 | 13.311 | 75.126 | 21.850 | 45.767 | 18.840 | 53.079 | 15.977 |
| $L/\Delta x = 100$ | 10,201 | 53.295 | 18.763 | 69.688 | 14.350 | 58.530 | 17.085 | 19.064 |
| $L/\Delta x = 200$ | 40,401 | 194.377 | 5.145 | 268.766 | 3.721 | 244.225 | 4.095 | 10.049 |

The computational efficiencies of this benchmark between the consistent $\delta^+$-ULPH model and the consistent $\delta^+$-SPH model are compared in Table 1. All cases in this paper are calculated by Intel(R) Core (TM) i7-14700K CPU with 64 GB RAM. The first 1,000 steps and three different particle resolutions with particle numbers ranging from 2,601 to 40,401 are considered. FPS represents the number of computational steps completed in a second. With the same searching range of neighboring particles (i.e., $h/\Delta x = 1.35$), the efficiency of the consistent $\delta^+$-ULPH model is slightly lower than that of the consistent $\delta^+$-SPH model (see Table 1). This is mainly caused by the extra computational cost due to the matrix operation in the governing equations. However, the accuracy of $\delta^+$-SPH with a small smoothing length is fairly low, and the above efficiency comparison is unfair. As a result, the efficiency of two models with similar accuracy (i.e., the consistent $\delta^+$-ULPH model with $h/\Delta x$ = 1.35 and the consistent $\delta^+$-SPH model with $h/\Delta x$ = 2.0) is compared. Speed-up rate is calculated by the FPS. Attributed to fewer neighboring particles, the consistent $\delta^+$-ULPH model has a speed-up rate between 10.049% to 19.064% (see Table 1). Since the computational devices and the quality of code writing play an important role in computational efficiency, the results are for reference only.

*4.2. Benchmark test No.2: Oscillating droplet*

A periodic oscillating droplet is a classical benchmark to verify the conservation and effectiveness of a new particle method [56,57]. A two-dimensional circular fluid is set up with radius $R$ = 1 m, initialized with the pressure and velocity fields written as [56]:

$$\begin{cases} \boldsymbol{u}_0(x,y) = \begin{pmatrix} \Omega_0 x & -\Omega_0 y \end{pmatrix}^T \\ p_0(x,y) = \frac{\rho_0 \Omega_0^2}{2}\left[R^2 - (x^2 + y^2)\right] \end{cases} \quad (35)$$

in which $\Omega_0$ is a constant and equals 1.0. In this benchmark, $\rho_0$ = 1.0 kg/m³. The droplet oscillates periodically, driven by a central conservative force field, written as:

$$\boldsymbol{F}(x,y) = \begin{pmatrix} -\Psi^2 x & -\Psi^2 y \end{pmatrix}^T \quad (36)$$

in which $\Psi$ is set to 1.0 in this benchmark. The corresponding oscillating period $T$ is 4.827 s. In this benchmark, $c_0 = 15\Omega_0 R$. A widely used $\alpha$ = 0.01 [53] is adopted.

To guarantee the accuracy of the tests, the convergence of the consistent $\delta^+$-ULPH model ($h/\Delta x$ = 1.35) is studied through the semi-axis $a(t)$ using three different particle resolutions ($R/\Delta x$ = 50, 100, and 200), as shown in Fig. 8. The analytical solution in the reference [58] is also plotted for comparison. It can be observed that the evolutions of dimensionless semi-axis $a(t)/R$ become more consistent when the particle resolution increases, and the results of $R/\Delta x$ = 100 and 200 are nearly identical, which indicates that the numerical results have converged. In addition, the results and analytical solution are in good agreement. In the following tests of this benchmark, the particle resolution $R/\Delta x = 100$ is adopted to balance the computational cost and accuracy.

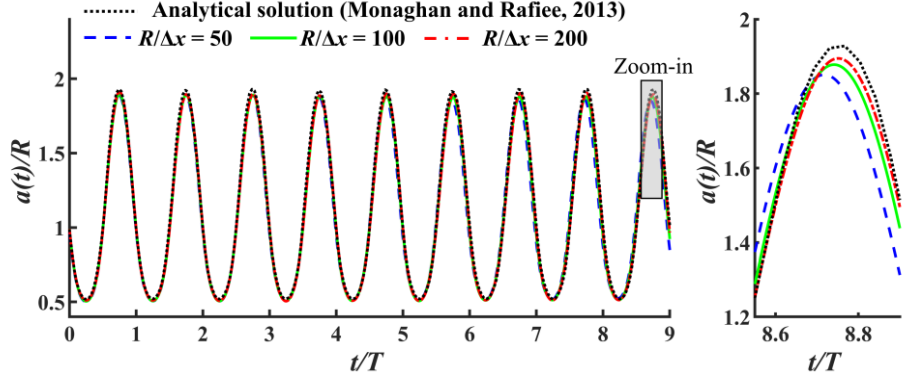

Fig. 8. Oscillating droplet: The horizontal semi-axis evolutions using the consistent $\delta^+$-ULPH model ($h/\Delta x = 1.35$) with three different resolutions ($R/\Delta x$ = 50, 100, and 200), compared with the analytical solution [58].

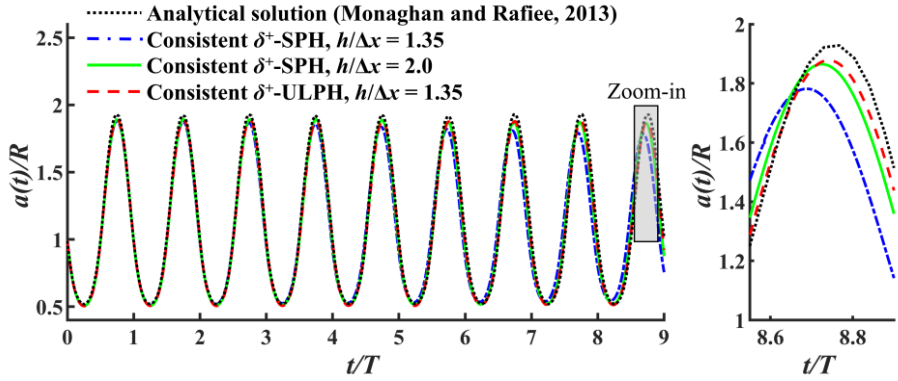

Fig. 9. Oscillating droplet: The horizontal semi-axis evolutions using the consistent $\delta^+$-ULPH model ($h/\Delta x = 1.35$, $R/\Delta x = 100$) and the consistent $\delta^+$-SPH model ($h/\Delta x = 1.35$ and 2.0, $R/\Delta x = 100$), compared with the analytical solution [58].

The evolutions of dimensionless semi-axis $a(t)/R$ calculated by the consistent $\delta^+$-ULPH and the consistent $\delta^+$-SPH models are shown in Fig. 9. The analytical solution in the reference [58] is also plotted. Results indicate that the consistent $\delta^+$-ULPH model with fewer neighboring particles (i.e., smaller smoothing length, $h/\Delta x = 1.35$) has similar accuracy to the consistent $\delta^+$-SPH model ($h/\Delta x = 2.0$), and both of them agree well with the analytical solution. However, when the number of neighboring particles decreases, the consistent $\delta^+$-SPH ($h/\Delta x = 1.35$) result is low precision, reflected in a smaller amplitude of $a(t)/R$ and a smaller period.

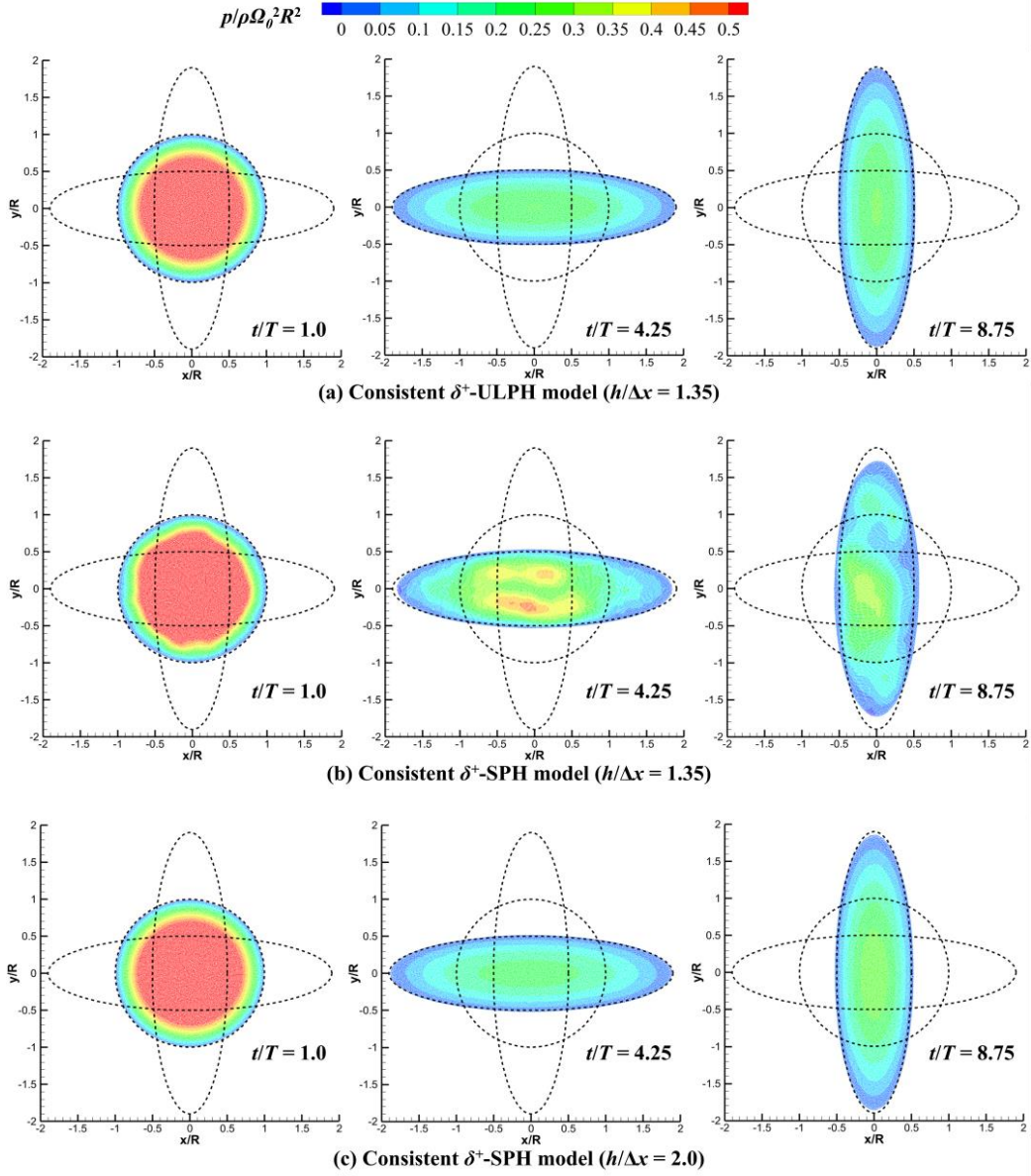

Fig. 10. Oscillating droplet: Pressure field comparison between the consistent $\delta^+$-ULPH ($h/\Delta x = 1.35$, $R/\Delta x = 100$) and the consistent $\delta^+$-SPH results ($h/\Delta x = 1.35$ and $2.0$, $R/\Delta x = 100$). The dashed line is the analytical solution [57] of the free-surface.

The pressure field evolutions of the consistent $\delta^+$-ULPH and the consistent $\delta^+$-SPH results are shown in Fig. 10. The analytical solution [57] of the free-surface is also plotted for comparison. The consistent $\delta^+$-ULPH model shows a remarkably stable pressure field, and its free-surface profile agrees well with the analytical solution, even in long-term simulation ($t/T = 8.75$, i.e., $t = 42.24$ s). Conversely, with the same particle resolution and smoothing length (i.e., $h/\Delta x = 1.35$), the consistent $\delta^+$-SPH model shows a worse pressure field due to numerical pressure waves, and its free-surface profile differs from the analytical solution, because the insufficient neighboring particles increase the discrete error. Since the acoustic damper is not used in the $\delta^+$-SPH model, pressure waves cannot be dissipated. When the smoothing length of the $\delta^+$-SPH scheme increases to $h = 2.0\Delta x$, the pressure field quality improves, and the free-surface profile approaches the analytical solution.

Energy conservation is investigated by analyzing the kinetic energy and mechanical energy evolutions, as shown in Fig. 11. To demonstrate the kinetic energy decay rate, the ratio of the kinetic energy $E_K/E_{K0}$ is utilized, where $E_{K0}$ and $E_K$ denote the initial and present kinetic energy, respectively. $E_K$ is calculated through the following formula:

$$E_K(t) = \sum_a^n \frac{1}{2} m_a \|\mathbf{u}_a(t)\|^2 \tag{37}$$

where $n$ is the total number of fluid particles. $m_a$ is the mass of particle $a$. Similarly, the ratio of the potential energy $E_P/E_{P0}$ is used to demonstrate the decay rate of the potential energy, where $E_{P0}$ and $E_P$ denote the initial and present potential energy, respectively. $E_P$ is calculated through the following formula:

$$E_P(t) = \sum_a^n \frac{1}{2} m_a \Omega_0 \left[ x_a^2(t) + y_a^2(t) \right] \tag{38}$$

In addition, the variable $\varepsilon_E$ is used to demonstrate the mechanical energy decay rate, expressed as follows:

$$\varepsilon_E(t) = \frac{\left| E_M(t) - E_{M_0} \right|}{E_M(t)} \times 100 \tag{39}$$

where $E_{M0}$ and $E_M$ are the initial and present mechanical energy, respectively.

As shown in Fig. 11, the energy decay of the consistent $\delta^+$-SPH model ($h/\Delta x = 1.35$) is fairly fast, and the mechanical energy decay rate reaches 10% when $t/T = 7.47$. With the value of $h/\Delta x$ increasing to 2.0, the energy decay rate decreases. Despite this, the decay rate of the consistent $\delta^+$-SPH model ($h/\Delta x = 2.0$) is still greater than that of the consistent $\delta^+$-ULPH model ($h/\Delta x = 1.35$). When $t/T = 9.0$, the mechanical energy of $\delta^+$-SPH is 1.69% lower than that of $\delta^+$-ULPH. The mechanical energy of the consistent $\delta^+$-

ULPH model slightly decays due to the artificial viscosity. In summary, results indicate that the present proposed ULPH scheme demonstrates lower numerical dissipation.

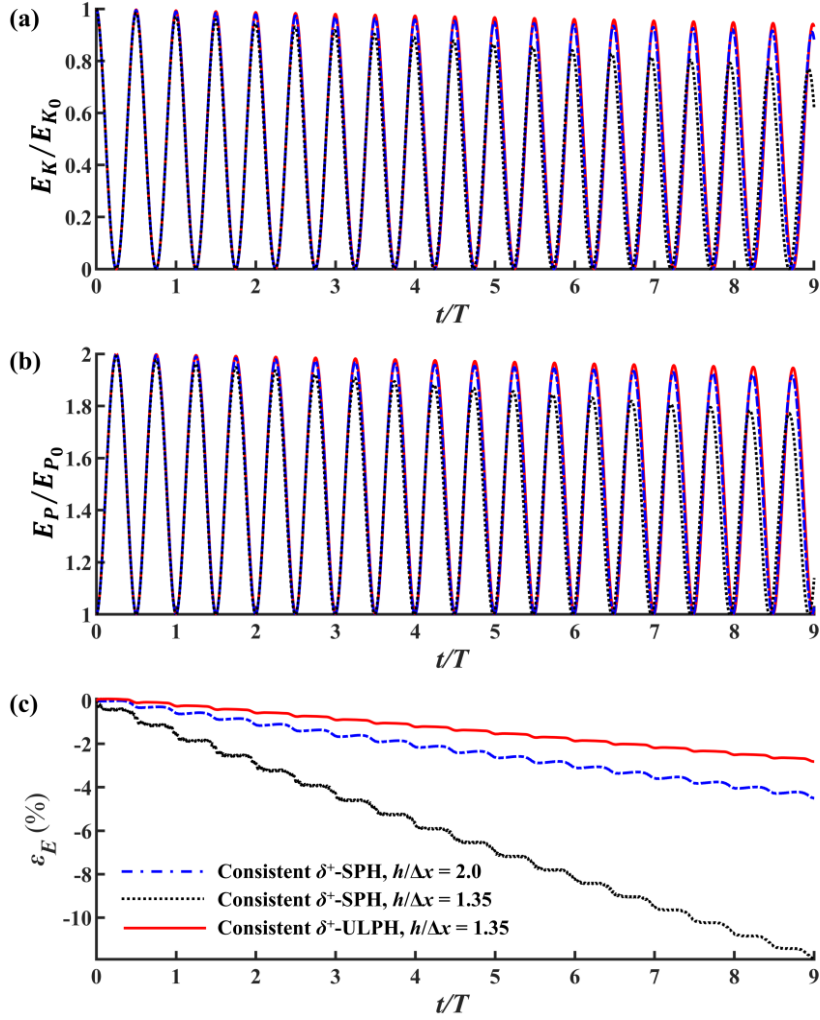

Fig. 11. Oscillating droplet: (a) Kinetic, (b) potential, and (c) mechanical energy decay evolutions using the consistent $\delta^+$-ULPH model ($h/\Delta x$ = 1.35, $R/\Delta x$ = 100) and the consistent $\delta^+$-SPH model ($h/\Delta x$ = 1.35 and 2.0, $R/\Delta x$ = 100). The artificial viscosity coefficient $\alpha$ = 0.01.

Table 2. Oscillating droplet: Computational efficiency comparison of the consistent $\delta^+$-ULPH and consistent $\delta^+$-SPH models. The computational cost is considered for the first 1,000 steps.

| Particle resolution | Particle number | The consistent $\delta^+$-SPH ($h/\Delta x$ = 2) | | The consistent $\delta+$-ULPH ($h/\Delta x$ = 1.35) | | |
|---|---|---|---|---|---|---|
| | | Computational cost (s) | FPS | Computational cost (s) | FPS | Speed-up rate (%) |
| $R/\Delta x$ = 50 | 7,668 | 54.609 | 18.312 | 47.873 | 20.889 | 14.071 |
| $R/\Delta x$ = 100 | 30,755 | 209.115 | 4.782 | 186.734 | 5.355 | 11.985 |
| $R/\Delta x$ = 200 | 123,119 | 795.493 | 1.257 | 674.854 | 1.482 | 17.876 |

With similar accuracy, the computational efficiencies of this benchmark between the consistent $\delta^+$-ULPH model ($h/\Delta x$ = 1.35) and the consistent $\delta^+$-SPH model ($h/\Delta x$ = 2.0) are compared in Table 2. The first 1,000 steps and three different particle resolutions with particle numbers ranging from 7,668 to 123,119 are considered. Attributed to fewer neighboring particles, the consistent $\delta^+$-ULPH model has a speed-up rate between 11.985% to 17.876%.

*4.3. Benchmark test No.3: Attenuation of a standing wave*

We are interested in whether the consistent $\delta^+$-ULPH model can avoid excessive numerical dissipation using fewer neighboring particles, especially in long-term and long-distance wave propagation simulations. A standing wave in long-term simulation is a classical benchmark widely used to investigate the energy conservation of a numerical model [25,27], with the advantage of relatively low computational cost and the characteristics of gravity waves in practical problems in coastal and ocean engineering. Standing waves with low and medium values of Reynolds number (i.e., Re = 500 and 2500) are simulated in this subsection to study the accuracy and energy conservation of the numerical models. In addition, an inviscid fluid is also considered.

A two-dimensional undisturbed water with length $L$ = 2 m and height $H$ = 1 m is set up, initialized with the pressure and velocity fields written as [25]:

$$\begin{cases} \boldsymbol{u}_0(x,y) = \begin{pmatrix} \varepsilon \dfrac{g\kappa H}{2\omega} \dfrac{\cosh(\kappa y)}{\cosh(\kappa H)} \sin\left[\kappa\left(x+\dfrac{L}{2}\right)\right] \\ -\varepsilon \dfrac{g\kappa H}{2\omega} \dfrac{\sinh(\kappa y)}{\cosh(\kappa H)} \cos\kappa\left[k\left(x+\dfrac{L}{2}\right)\right] \end{pmatrix} \\ p_0(x,y) = \rho g(H-y) \end{cases} \quad (40)$$

in which $\varepsilon = 2A/H$ denotes the nonlinearity parameter representing wave steepness and equals 0.1 in this paper, where $A$ is the wave amplitude and equals 0.05 m. $\kappa = 2\pi/\lambda_0$ denotes the wave number, where $\lambda_0$ is the wavelength, which equals 2.0 m in this benchmark. $\omega$ is the angular frequency calculated by $\omega^2 = g\kappa\tanh(\kappa H)$. Correspondingly, the oscillation period $T$ of the standing wave equals 1.1339 s. In this benchmark, $\rho_0$ = 1000 kg/m³ and $c_0 = 10\sqrt{gH}$.

The boundary conditions of this benchmark are as follows. The periodic boundary [27] is applied to the lateral boundaries, which allows the fluid particles to freely move across the computational domain from one side to the other without any extra damping effects. To ensure the impenetrable condition, the mirror boundary [59] is applied to the bottom, as a result of its fairly high accuracy and convenient implementation. More details about the mirror boundary can be referred in [59].

The ratio $E_K/E_{K0}$ is used to demonstrate the kinetic energy decay rate. The analytical solution of the kinetic energy decay of the standing wave is written as [60]:

$$E_K(t) = \varepsilon^2 g \frac{\lambda_0 H^2}{32} e^{-4\kappa^2 \nu t} \left[1 + \cos(2\omega t)\right] \quad (41)$$

in which $\nu$ represents the kinematic viscosity. The relationship between Reynolds number and kinematic viscosity is written as follows:

$$\mathrm{Re}(\nu) = H\sqrt{gH}/\nu \quad (42)$$

By comparing the $E_K/E_{K0}$ evolutions, the convergence of the consistent $\delta^+$-ULPH model ($h/\Delta x = 1.35$) is studied with three different particle resolutions ($H/\Delta x = 50, 100,$ and 200), as shown in Fig. 12. The analytical solutions are also plotted for comparison [60]. It can be observed that the time histories of $E_K/E_{K0}$ become more consistent when the particle resolution increases, and the results of $H/\Delta x = 100$ and 200 are nearly identical and agree well with the analytical solution. This indicates that the numerical results have converged. In the following tests of this benchmark, the particle resolution $H/\Delta x = 100$ is adopted to balance the computational cost and accuracy.

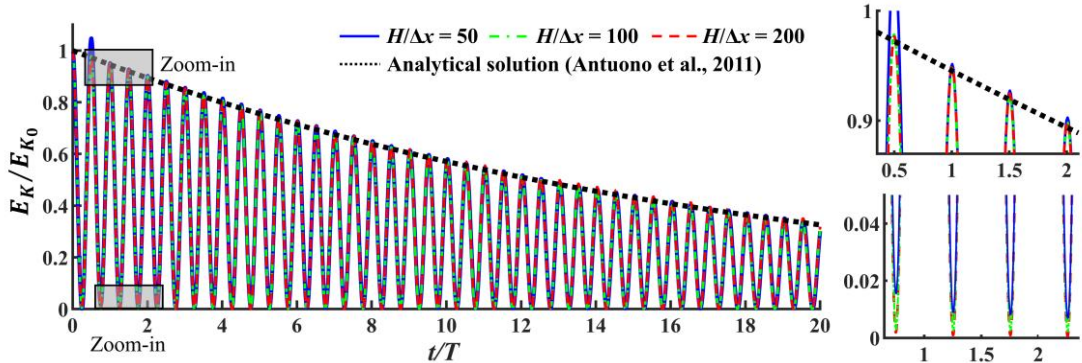

Fig. 12. Standing wave with Re = 2500: Kinetic energy decay using the consistent $\delta^+$-ULPH model ($h/\Delta x = 1.35$) with three different resolutions ($H/\Delta x = 50, 100,$ and 200), compared with the analytical solution [60].

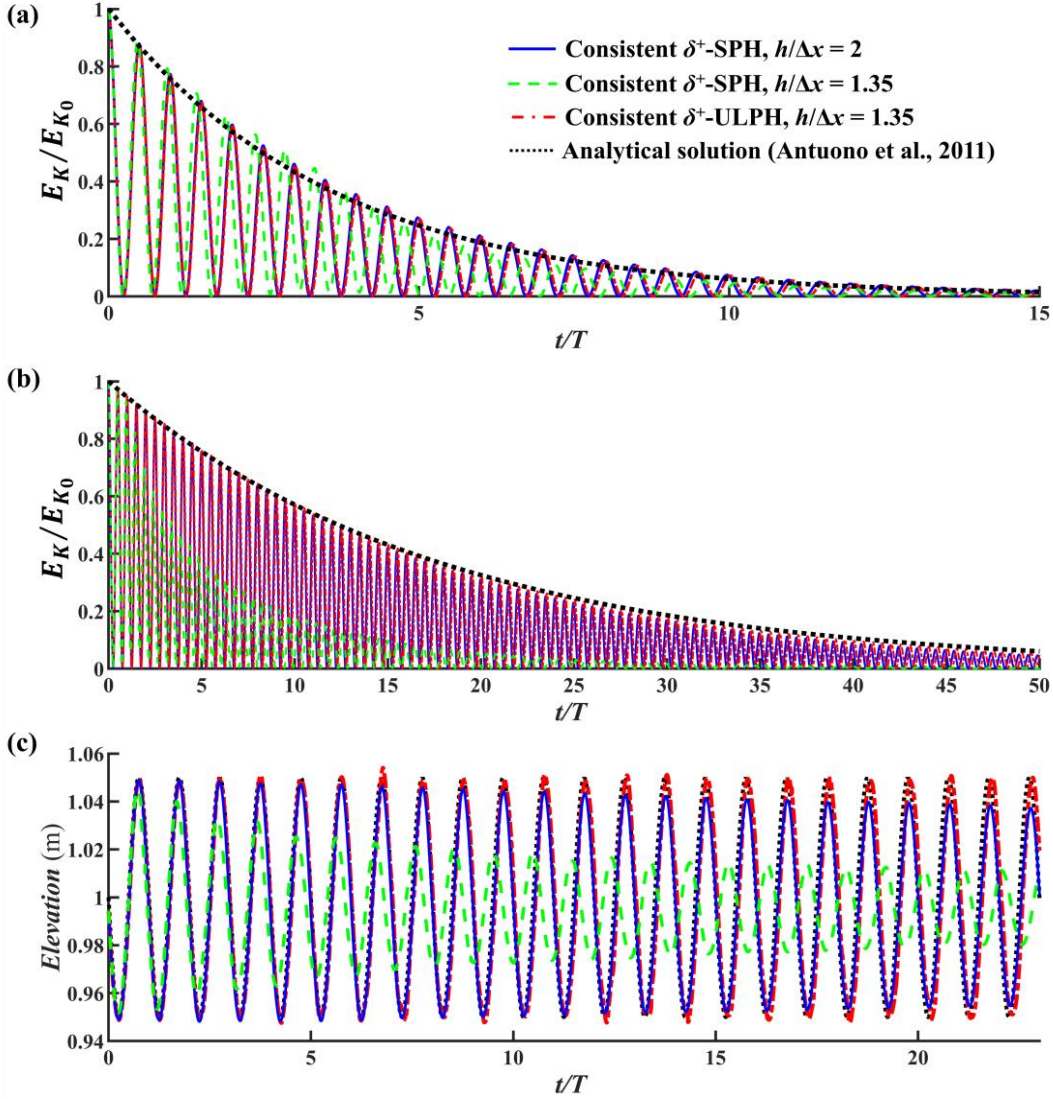

Fig. 13. Standing wave: Comparison of the numerical dissipation between the consistent $\delta^+$-ULPH ($h/\Delta x$ = 1.35, $H/\Delta x$ = 100) and the consistent $\delta^+$-SPH results ($h/\Delta x$ = 1.35 and 2.0, $H/\Delta x$ = 100). (a) The kinetic energy decay with Re = 500, (b) the kinetic energy decay with Re = 2500, and (c) the wave elevation at $x$ = 0 with an inviscid fluid.

Figure 13 shows the numerical dissipation of the consistent $\delta^+$-ULPH and the consistent $\delta^+$-SPH models by comparing the kinetic energy decay (viscous fluid with Re = 500 and 2500) and wave elevation (inviscid fluid). It can be noticed that even with fewer neighboring particles ($h/\Delta x$ = 1.35), the consistent $\delta^+$-ULPH results show fairly low numerical dissipation and have a good agreement with the analytic solutions, while the excessive dissipation of the consistent $\delta^+$-SPH model ($h/\Delta x$ = 1.35) is visible and further intensifies with the fluid viscosity decreasing. Especially in the inviscid fluid case, when $t/T$ = 20, the elevation amplitude decay of $\delta^+$-ULPH is negligible, while the amplitude of $\delta^+$-SPH is 76.4% lower than the analytical solution due to the excessive dissipation. The dissipation of $\delta^+$-SPH reduces when the value of $h/\Delta x$ increases to 2.0. Despite this, the energy conservation of the consistent $\delta^+$-SPH model is still weaker than that of the consistent $\delta^+$-ULPH model.

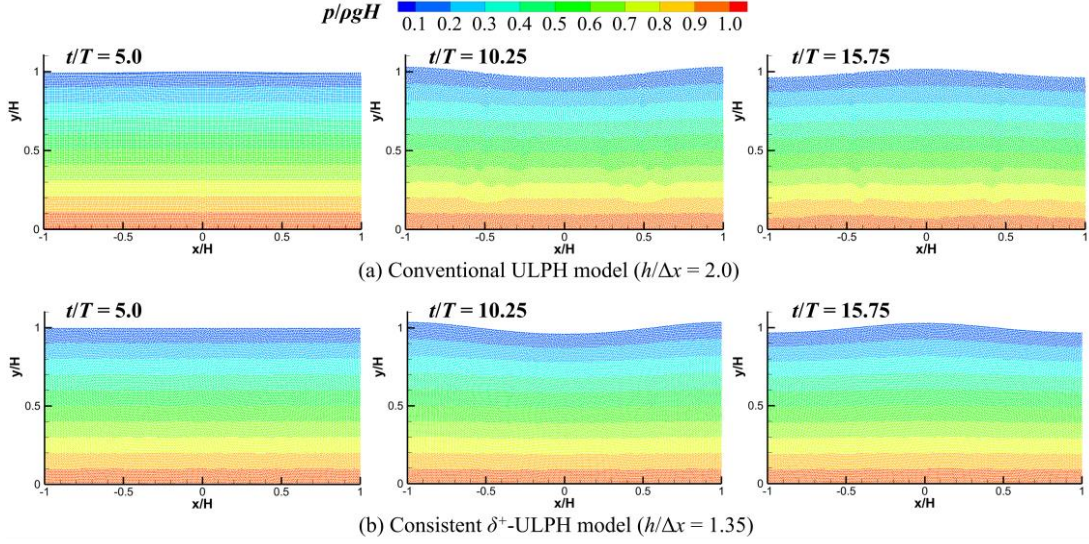

Fig. 14. Standing wave simulated with Re = 2500: Pressure field comparison between (a) the conventional ULPH without optimal matrix **D**, and (b) the consistent $\delta^+$-ULPH results ($H/\Delta x = 100$).

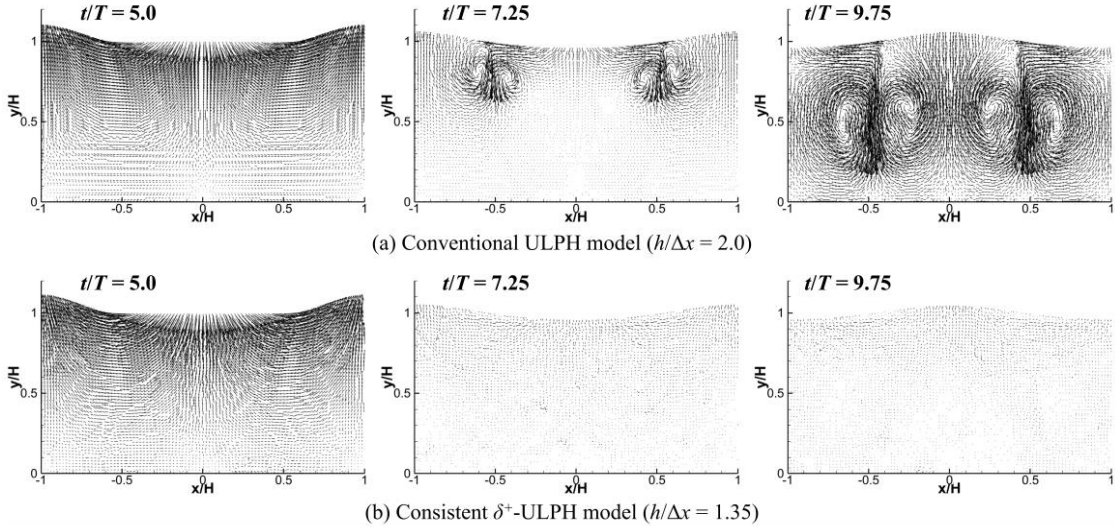

Fig. 15. Standing wave simulated with Re = 2500: Comparison of the velocity field evolutions between (a) the conventional ULPH without optimal matrix **D**, and (b) the consistent $\delta^+$-ULPH results ($H/\Delta x = 100$).

Table 3. Standing wave simulated with Re = 2500: Computational efficiency comparison of the consistent $\delta^+$-ULPH and consistent $\delta^+$-SPH models. The computational cost is considered for the first 1,000 steps.

| Particle resolution | Particle number (without boundary) | The consistent $\delta^+$-SPH ($h/\Delta x = 2$) | | The consistent $\delta$+-ULPH ($h/\Delta x = 1.35$) | | |
|---|---|---|---|---|---|---|
| | | Computational cost (s) | FPS | Computational cost (s) | FPS | Speed-up rate (%) |
| $H/\Delta x = 50$ | 5,000 | 60.198 | 16.612 | 39.234 | 25.488 | 53.433 |
| $H/\Delta x = 100$ | 20,000 | 191.089 | 5.233 | 129.661 | 7.712 | 47.376 |
| $H/\Delta x = 200$ | 80,000 | 656.413 | 1.523 | 475.275 | 2.104 | 38.112 |

As shown in Figs. 14 and 15, the pressure and velocity field evolutions obtained by the conventional ULPH without optimal matrix $\boldsymbol{D}$ and the consistent $\delta^+$-ULPH models are compared. The pressure field of the consistent $\delta^+$-ULPH model is fairly good. Conversely, the pressure field of the conventional ULPH model becomes worse in long-term simulation, due to the free-surface instability. As shown in Fig. 15, in conventional ULPH simulation, approximately after the seventh period, the free-surface becomes unstable owing to the kernel function truncation, characterized by unphysical vortices generating and growing up, while the free-surface of the consistent $\delta^+$-ULPH model consistently maintains good stability, attributed to superior conservation from ESDT and the optimal matrix $\boldsymbol{D}$ proposed in Section 3.3.

With similar accuracy, the computational efficiencies of this benchmark between the consistent $\delta^+$-ULPH model ($h/\Delta x = 1.35$) and the consistent $\delta^+$-SPH model ($h/\Delta x = 2.0$) are compared in Table 3. The first 1,000 steps and three different particle resolutions with particle numbers (without boundary particles) ranging from 5,000 to 80,000 are considered. Attributed to fewer neighboring particles, the consistent $\delta^+$-ULPH model has a speed-up rate between 38.112% to 53.433%. The speed-up rate of this benchmark is visibly greater than that of benchmarks No.1 and No.2 (see Tables 1 and 2), since the lower proportion of the free-surface. According to Eq. (27), when calculating the smallest distance $l_i$ between target particle $i$ and the free-surface particles, a lower free-surface proportion causes lower extra computational cost.

*4.4. Benchmark test No.4: Long-distance propagation of a regular wave*

Apart from the long-term standing wave benchmark, a long-distance propagation of a regular wave is simulated to verify the low numerical dissipation of the consistent $\delta^+$-ULPH model in this subsection. The numerical setup is consistent with the experiment conducted by Huang et al. [36]. A two-dimensional numerical wave tank is set up with length $L = 20$ m, height $H = 0.5$ m, and water depth $d = 0.266$ m, as shown in Fig. 16. Three wave gauges (WGs) are placed on $x = 2.37$, 6.37, and 15.0 m along the length direction of the wave tank to measure the wave elevations. In this case, $\rho_0 = 1000$ kg/m$^3$ and $c_0 = 20.0$ m/s. The $\alpha = 0.02$ suggested by Huang et al. [36] is adopted.

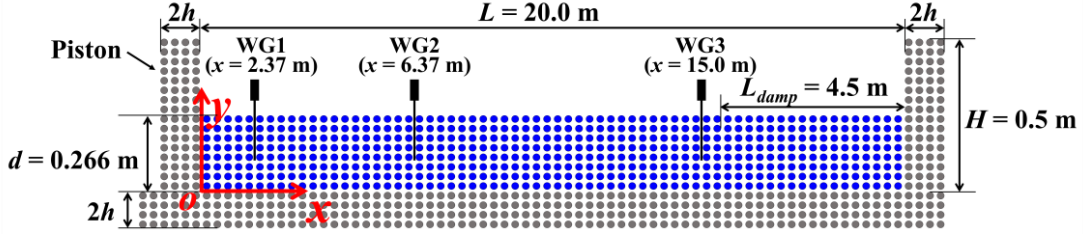

Fig. 16. Propagation of a regular wave: Sketch of a wave tank.

A wave generation with a piston moving along the *x*-direction is equipped on the left side of the tank. The experimental data of the piston motion in reference [36] is input to drive the numerical piston. A wave absorbing zone with length $L_{\text{damp}} = 4.5$ m is set up on the other side, where the artificial viscosity coefficient $\alpha$ that linearly increases along the positive *x*-direction instead of a constant value of 0.02 is applied to achieve the wave attenuation. The $\alpha$ is defined as follows:

$$\alpha = \begin{cases} 0.02, & 0 < x < L - L_{\text{damp}} \\ 1.0\left[x - \left(L - L_{\text{damp}}\right)\right]/L_{\text{damp}}, & L - L_{\text{damp}} \leq x < L \end{cases} \tag{43}$$

Ghost particle boundary [61] is applied to avoid wall penetration. The primary procedure is to calculate the pressure of ghost wall particles by interpolation using their neighboring fluid particles. More details can be found in reference [61]. It is worth noticing that at fluid-solid interface, the calculation of several terms involving density should be adjusted, since the density of the ghost solid particles is spurious and may cause numerical instability [62,63]. Considering a fluid particle *i*, when its neighboring particle *j* is within the solid phase, the contributions of particle *j* in the $\delta \boldsymbol{u}$-terms in Eq. (29) and the density diffusive term $\Phi_i$ in Eq. (16) should not be considered. Additionally, by setting the viscous force between fluid and ghost wall particles to zero, the free-slip boundary is applied.

Through comparing the wave elevations $\eta_1$ and $\eta_2$ (recorded by WG1 and WG2, respectively), the convergence of the consistent $\delta^+$-ULPH model ($h/\Delta x = 1.35$) is studied with three different particle resolutions ($d/\Delta x = 15$, 30, and 60), as shown in Fig. 17. The experimental results [36] are also plotted for validation. It can be observed that when the particle resolution increases, the wave elevations become more consistent with the experiment, and the results of $d/\Delta x = 60$ agree well with the experiment. Additionally, the pressure field (see Fig. 18) obtained by the consistent $\delta^+$-ULPH model with particle resolution $d/\Delta x = 60$ is fairly good. Consequently, the particle resolution $d/\Delta x = 60$ is adopted in the following tests of this benchmark.

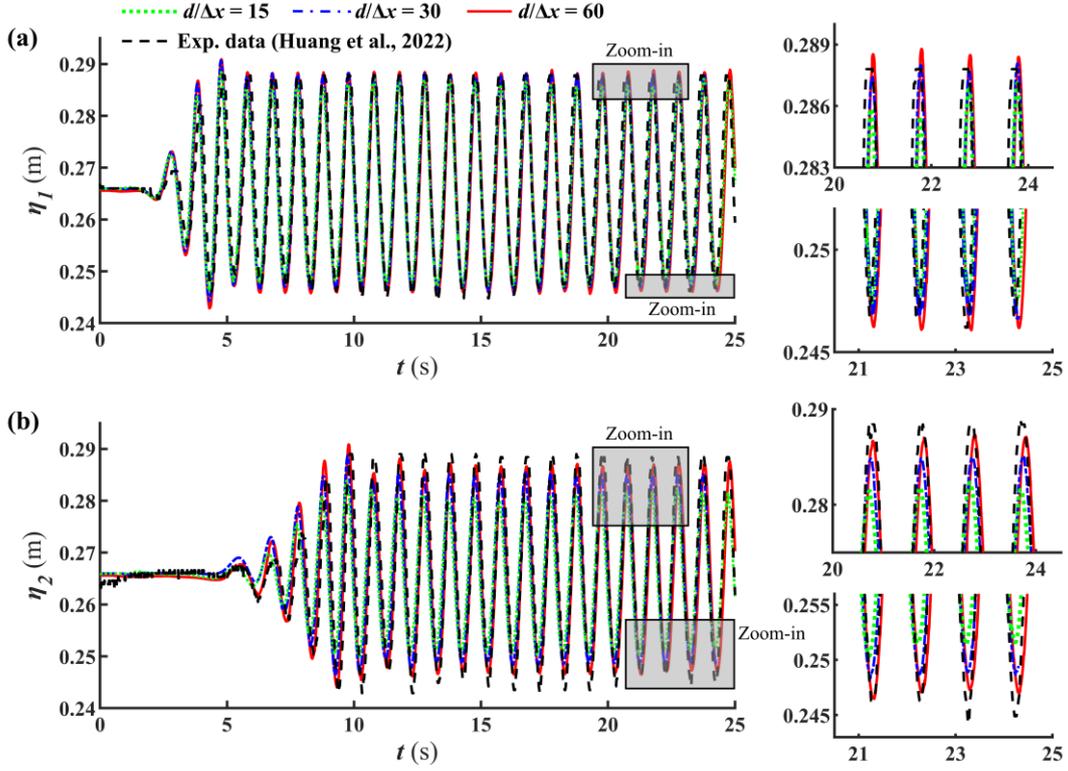

Fig. 17. Propagation of a regular wave: Free-surface elevations at (a) $x = 2.37$ m and (b) $x = 6.37$ m using the consistent $\delta^+$-ULPH model ($h/\Delta x = 1.35$) with three different resolutions ($d/\Delta x = 15$, 30, and 60), compared with the experimental results [36].

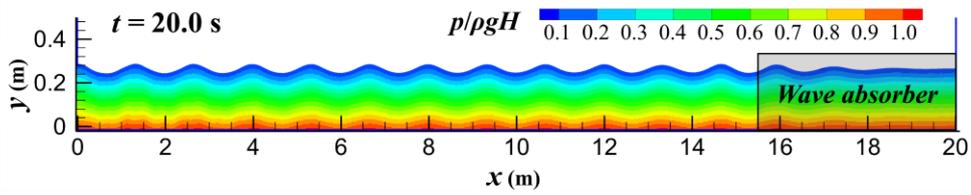

Fig. 18. Propagation of a regular wave: Pressure field obtained by the consistent $\delta^+$-ULPH model ($h/\Delta x = 1.35$, $d/\Delta x = 60$).

Figure 19 shows the wave elevations of the consistent $\delta^+$-ULPH and the consistent $\delta^+$-SPH results. Results show that $\eta_1$ and $\eta_2$ of the consistent $\delta^+$-ULPH model ($h/\Delta x = 1.35$) agree well with the experiment [36], while the amplitudes of $\eta_1$ and $\eta_2$ obtained by the consistent $\delta^+$-SPH model ($h/\Delta x = 1.35$) are approximately 18.6% and 40.9% lower than those of the experiment, respectively, mainly due to the excessive numerical dissipation. When the number of neighboring particles increases, $\eta_1$ of the $\delta^+$-SPH model ($h/\Delta x = 2.0$) matches well with the experiment. However, as the wave propagates, $\eta_2$ is still slightly lower than that of the experiment. With the wave propagating farther, the wave amplitudes of $\eta_3$ (recorded by WG3) obtained by both numerical models are lower than the initial one, attributed to the artificial viscosity and numerical dissipation. Despite this, the dissipation of the consistent $\delta^+$-ULPH model is still visibly lower than that of the consistent $\delta^+$-SPH model.

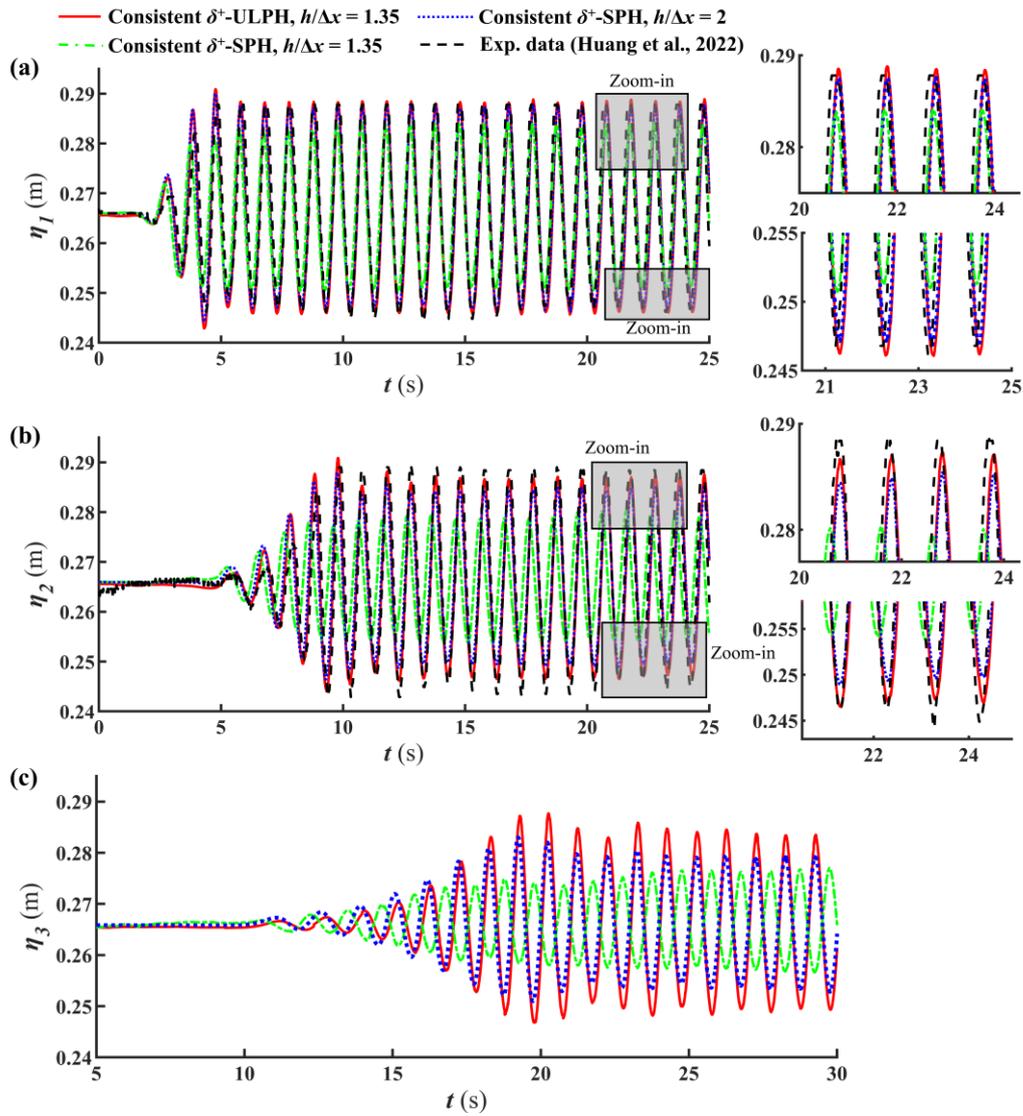

Fig. 19. Propagation of a regular wave: Free-surface elevations at (a) $x = 2.37$ m, (b) $x = 6.37$ m, and (c) $x = 15$ m using the consistent $\delta^+$-ULPH model ($h/\Delta x = 1.35$, $d/\Delta x = 60$) and the consistent $\delta^+$-SPH model ($h/\Delta x = 1.35$ and 2.0, $d/\Delta x = 60$), compared with the experimental results [36].

With similar accuracy, the computational efficiencies of this benchmark between the consistent $\delta^+$-ULPH model ($h/\Delta x = 1.35$) and the consistent $\delta^+$-SPH model ($h/\Delta x = 2.0$) are compared in Table 4. The first 1,000 steps and three different particle resolutions with particle numbers ranging from 22,840 to 289,720 are considered. Attributed to fewer neighboring particles, the consistent $\delta^+$-ULPH model has a speed-up rate between 42.545% to 55.725%. Since the lower proportion of the free-surface, the speed-up rate of this benchmark is also visibly greater than that of benchmarks No.1 and No.2 (see Tables 1 and 2).

Table 4. Propagation of a regular wave: Computational efficiency comparison of the consistent $\delta^+$-ULPH and consistent $\delta^+$-SPH models. The computational cost is considered for the first 1,000 steps.

| Particle resolution | Particle number | The consistent $\delta^+$-SPH ($h/\Delta x = 2$) | | The consistent $\delta+$-ULPH ($h/\Delta x = 1.35$) | | |
|---|---|---|---|---|---|---|
| | | Computational cost (s) | FPS | Computational cost (s) | FPS | Speed-up rate (%) |
| $d/\Delta x = 15$ | 22,840 | 191.945 | 5.210 | 123.259 | 8.113 | 55.725 |
| $d/\Delta x = 30$ | 79,464 | 608.812 | 1.643 | 427.103 | 2.341 | 42.545 |
| $d/\Delta x = 60$ | 289,720 | 2021.698 | 0.495 | 1394.765 | 0.717 | 44.949 |

*4.5. Benchmark test No.5: Violent vertical sloshing*

In the final benchmark, violent liquid sloshing in a vertically oscillating rectangular tank is simulated and compared to the experiment [64-66] to validate the consistent $\delta^+$-ULPH model. It is a challenging benchmark with high-frequency acceleration, violent impacts, highly fragmented free-surface, turbulence, and fluid-structure interaction.

Consistent with the experiments, a two-dimensional sloshing tank filled with 50% liquid is set up (see Fig. 20). The detailed parameters are listed in Table 5. Only a single DoF (i.e., surge motion) is allowed. The tank is connected with an upper and a lower set of springs in experiments, while in numerical simulations, elastic potential energy is directly imposed on the tank according to the total stiffness and deformation of the springs to simplify the physical model, instead of modeling the real springs. As a result, the spring mass is neglected and the total mass of the system is $m_{total} = m_s + m_l = 2.222$ kg. The analytical oscillation frequency is defined as $f_0 = \sqrt{K_{Spring}/m_{total}}/(2\pi) = 7.018$ Hz. The oscillation period is $T = 1/f_0 = 0.142$ s. The circular frequency is $\omega = 2\pi f_0 = 44.093$ rad/s. Initially, the y-position of the tank is 0, and the maximum spring deformation relative to position in steady state is $y_0 = 0.059$ m. $c_0 = 40.0$ m/s and $K_c = 0.8$ in this benchmark.

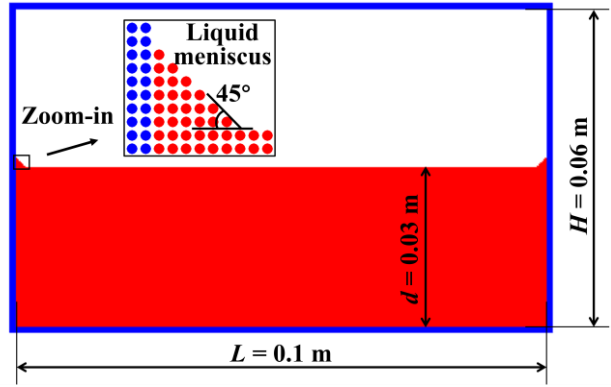

Fig. 20. Violent vertical sloshing: Sketch of sloshing tank.

Table 5. The detailed parameters of the sloshing system.

| Tank | | | | Liquid (Oil) | | | Spring |
|---|---|---|---|---|---|---|---|
| Length $L$ (m) | Height $H$ (m) | Depth $d$ (m) | Mass $m_s$ (kg) | Density $\rho_0$ (kg/m³) | Mass $m_l$ (kg) | Dynamic viscosity $\mu$ (Pa·s) | Stiffness $K_{Spring}$ (N/m) |
| 0.1 | 0.06 | 0.03 | 2.06 | 900 | 0.162 | 0.045 | 4320 |

Several assumptions are as follows. According to Marrone et al. [65], the single-phase assumption is sufficiently accurate when evaluating the sloshing dissipation in this benchmark. Malan et al. [67] compared the sloshing results obtained by a single-phase SPH and two-phase Finite Volume Method. The vertical sloshing dissipation between the two models is quite similar, which means the role of the air phase is not critical in the slosh dissipation. Considering the length and velocity scales, neglecting the surface tension does not significantly affect the accuracy of the sloshing dissipation, as reported in reference [66]. Additionally, the single-phase SPH models without surface tension are also used to study this vertical sloshing benchmark in references [68,69], and the single-phase numerical results agree well with the experimental data. Consequently, the air phase and surface tension are neglected.

To better compare the evolution of free-surface deformation between the ULPH results and the experimental snapshots at the initial stage, two liquid menisci (see Fig. 20) are initially set up at the corner between the liquid and the walls as suggested by Marrone et al. [69]. The liquid meniscus is 1.5 mm height with a 45° angle, which physically exists due to the contact angle of the fluid on the wall boundary. These liquid menisci play an important role in the accurate development of the liquid surface deformation at the initial sloshing stage. Relevant investigations about the effects of liquid meniscus have been shown in reference [66].

Ghost particle boundary [61] is employed on the tank wall. Additionally, an effective anti-penetration technique [70] between the fluid and ghost particles is adopted. The basic procedure is, when a fluid particle approaches the ghost particle boundary and is detected as a particle with penetration potential, its normal velocity component is replaced by that of its nearest ghost particle. More details can be referred in [70]. Additionally, physical viscosity and a no-slip boundary are used in this benchmark.

As reported in reference [64], the vertical motion of the sloshing tank follows the following equation, which is derived from Newton's second law, written as:

$$m_s \frac{d^2 y}{dt^2} = F_y^{slosh} - B_{0d} \cdot sign\left(\frac{dy}{dt}\right) - B_{1d} \frac{dy}{dt} - Ky \tag{44}$$

in which $B_{0d} = 0.38$ N and $B_{1d} = 1.73$ kg/s denote the Coulomb damping coefficient and the viscous damping coefficient, respectively. $F_y^{slosh}$ denotes the fluid-structure interaction force $\boldsymbol{F}_{f-s}$ in the y-direction. $\boldsymbol{F}_{f-s}$ is calculated by:

$$\boldsymbol{F}_{f-s} = \sum_{i \in fluid} \sum_{j \in solid} \left[ W_{ij} \left( p_i \boldsymbol{M}_i^{-1} + p_j \boldsymbol{M}_j^{-1} \right) \boldsymbol{r}_{ji} V_j - \boldsymbol{F}_i^V \right] V_i \tag{45}$$

in which $\boldsymbol{F}_i^V$ is the viscosity term obtained by Eq.(17). It's worth noting that when applying the no-slip boundary, the boundary velocity participating in the calculation of $\boldsymbol{F}_i^V$ is obtained from interpolation. More details can be referred in [61].

A convergence study of the consistent $\delta^+$-ULPH model ($h/\Delta x = 1.35$) is conducted with three different particle resolutions ($d/\Delta x = 25$, 50, and 100), as shown in Fig. 21. The experimental results [65,66] are also plotted for validation. The computed sloshing forces are filtered through a cutoff frequency of 50 Hz, consistent with the experiment [66]. As the particle resolution increases, the numerical results become more consistent with the experiment, and the results of $d/\Delta x = 100$ agree well with the experiment. Consequently, in the following tests of this benchmark, the particle resolution $d/\Delta x = 100$ is adopted.

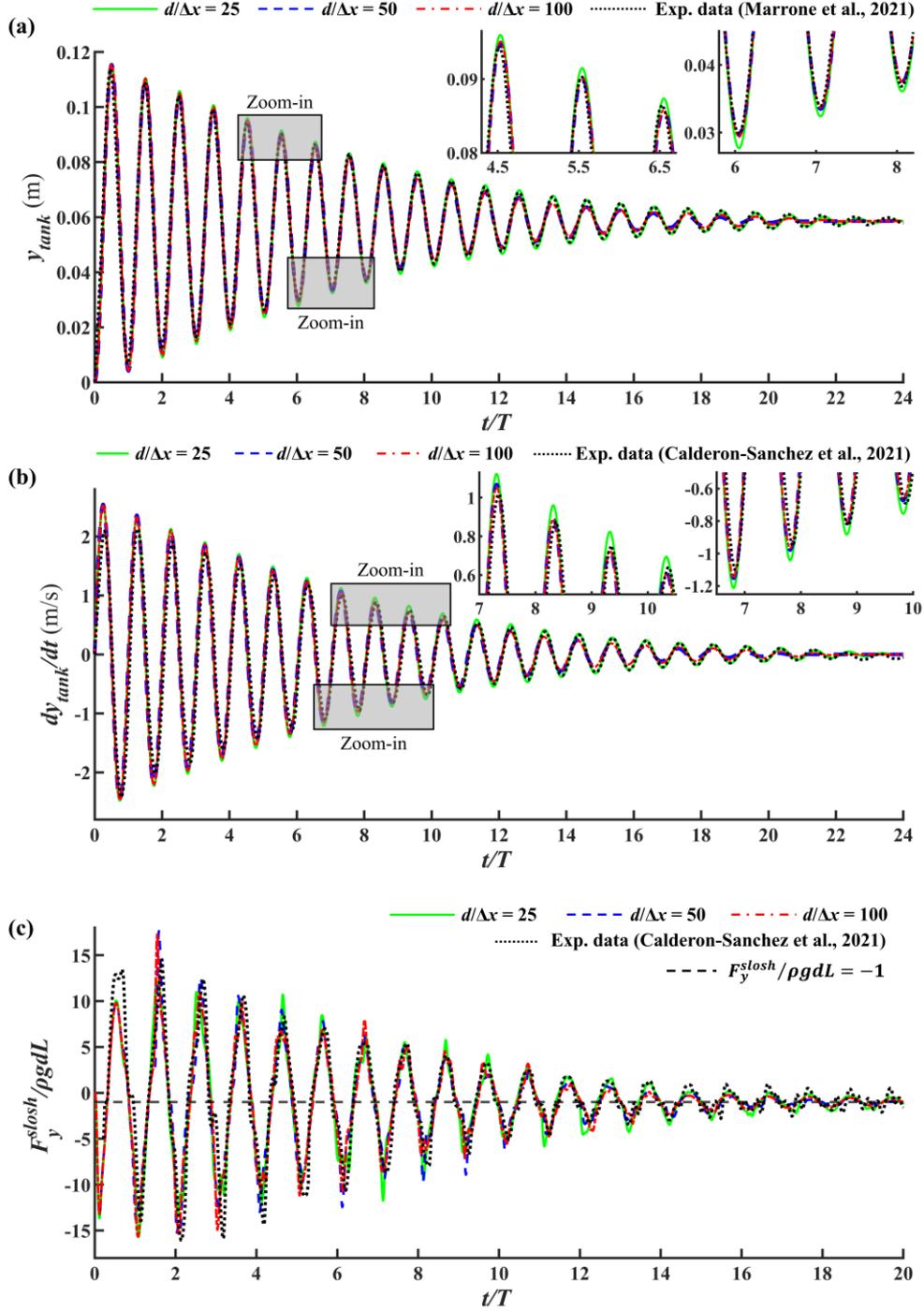

Fig. 21. Violent vertical sloshing: Time history of (a) tank displacement $y_{\text{tank}}$, (b) tank velocity $dy_{\text{tank}}/dt$, and (c) sloshing force in the y-direction using the consistent $\delta^+$-ULPH model ($h/\Delta x = 1.35$) with three different resolutions ($d/\Delta x = 25$, 50, and 100), compared with experimental data [65,66].

The flow evolutions of the consistent $\delta^+$-ULPH model and experiment [64,65] are shown in Fig. 22. The numerical results agree well with the experiment. The flow field evolutions can be described through several stages. First, when the tank initially accelerates upwards, the menisci rupture, disturbing the free-surface stability and leading to two tiny free-surface waves that generate and travel from the wall sides towards the tank center. When the acceleration direction of the tank changes for the first time, the Rayleigh-Taylor instability (see Fig. 22, $t/T = 0.45$ and $0.53$) arises as a result of these disturbances. Since the tank has the opposite acceleration to the fluid, the negative pressure generates and may lead to tensile instability and numerical cavitation [69] when using the conventional SPH and ULPH models. While the consistent $\delta^+$-ULPH model effectively avoids this problem attributed to the PST and TIC. As the fluid continues to accelerate upwards, two primary jets generate and first impact against the tank ceiling (see Fig. 22, $t/T = 0.70$ and $0.80$). The major stage of the vertical sloshing is characterized by chaos and turbulence (see Fig. 22, $t/T = 5.19$ and $8.98$). Eventually, when the tank motion is fairly slight, a standing-wave-like behavior develops (see Fig. 22, $t/T = 23.74$ and $26.77$).

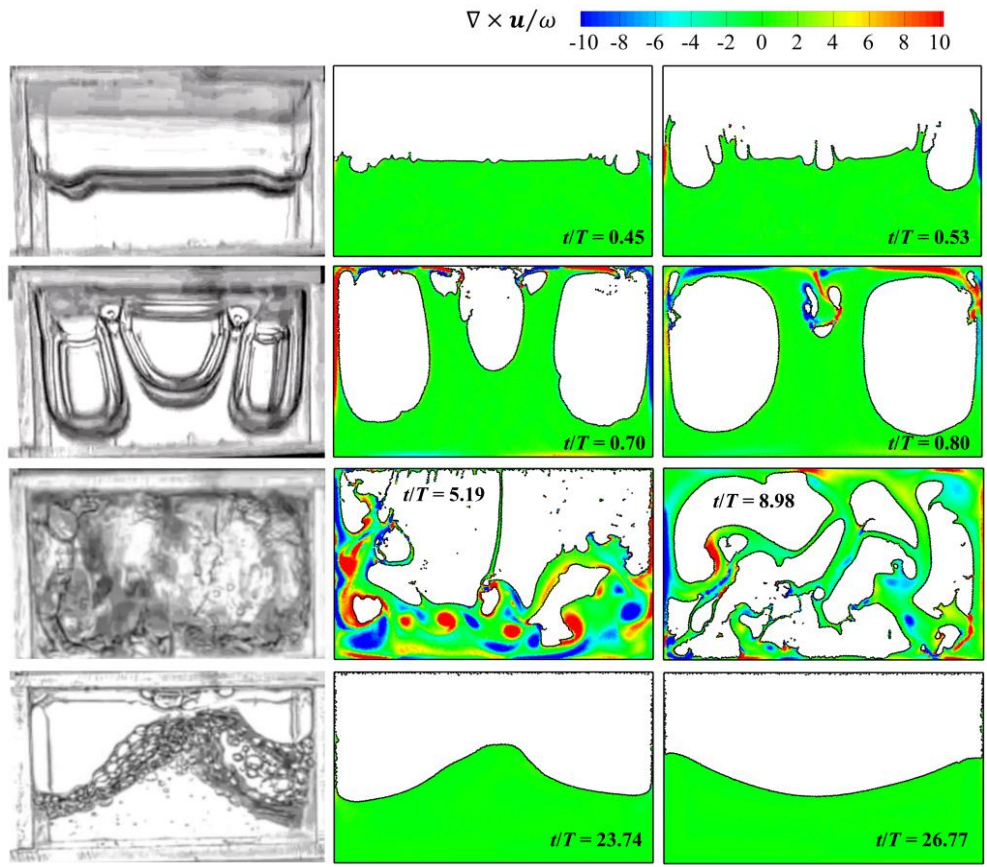

Fig. 22. Violent vertical sloshing: Comparison of the flow evolutions between the consistent $\delta^+$-ULPH result ($h/\Delta x = 1.35$, $d/\Delta x = 100$) and experiment [64,65]. The numerical vorticity is displayed.

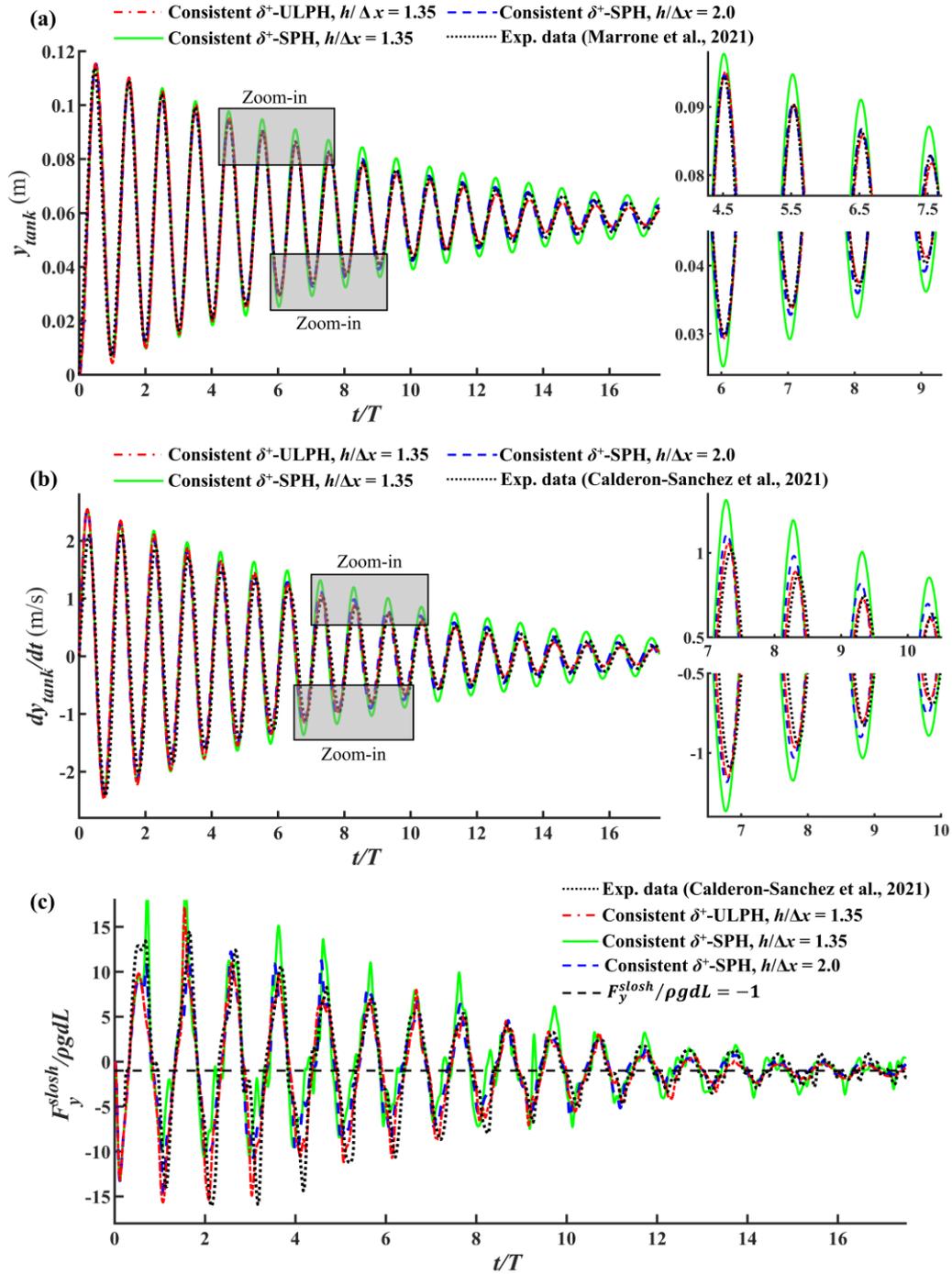

Fig. 23. Violent vertical sloshing: Time history of (a) tank displacement $y_{tank}$, (b) tank velocity $dy_{tank}/dt$, and (c) sloshing force in y-direction using the consistent $\delta^+$-ULPH model ($h/\Delta x = 1.35$, $d/\Delta x = 100$) and the consistent $\delta^+$-SPH model ($h/\Delta x = 1.35$ and 2.0, $d/\Delta x = 100$), compared with the experimental results [65,66].

Figure 23 compares the numerical results of the consistent $\delta^+$-ULPH model and the consistent $\delta^+$-SPH model. The experimental results [65,66] are also plotted for validation. It can be noticed that the consistent $\delta^+$-ULPH results ($h/\Delta x = 1.35$) agree well with the experimental results, while visible differences can be observed between the consistent $\delta^+$-SPH results ($h/\Delta x$ = 1.35) and the experiments. Increasing the neighboring particle number improves the accuracy of the $\delta^+$-SPH results ($h/\Delta x$ = 2.0). However, several peaks of the sloshing force (see Fig. 23 (c), 1.5 < $t/T$ < 4.5) acting on the bottom of the tank are still difficult to accurately capture by the consistent $\delta^+$-SPH model. This problem of the SPH results is also displayed in reference [66].

The volume conservation of the present model is analyzed, since it is an important issue of the weakly-compressible SPH model in violent free-surface flows characterized by strong fragmentation [68]. In the study on conservation, the TIC is not applied since it weakens the conservation property.

The volume error $\varepsilon_V$ is calculated by the following formula [68]:

$$\varepsilon_V = \frac{\sum_i V_i}{V_{\text{int}}} - 1, \quad V_i = \begin{cases} (\Delta x)^2 & \text{if } i \in (F \bigcup V) \\ 1/\sum_j W_{ij} & \text{else} \end{cases} \quad (46)$$

in which $V_{\text{int}}$ is the initial volume. $V_i$ is the volume of particle $i$. $\Delta x$ represents the initial particle distance. To obtain superior volume conservation, an enforcement of dynamic boundary conditions (EdBC) correction [68] is adopted, expressed as follows:

$$p_i = 0, \quad \rho_i = \rho_0, \quad V_i = m_i/\rho_0, \quad \text{if } i \in (F \bigcup V), \lambda_i < 0.2, \text{ and } \varepsilon_V < 0 \quad (47)$$

in which $\lambda_i$ is the minimum eigenvalue of the detection matrix $M_i^D$ in Eq. (8), and $\lambda_i$ < 0.2 represents the particle of thin jets or drops. Since the majority of the volume errors are caused by the free-surface fragmentation, the EdBC correction has a significant effect. The present smoothing length $h$ = 2$\Delta x$. For different smoothing lengths and kernel functions, the value 0.2 can be altered.

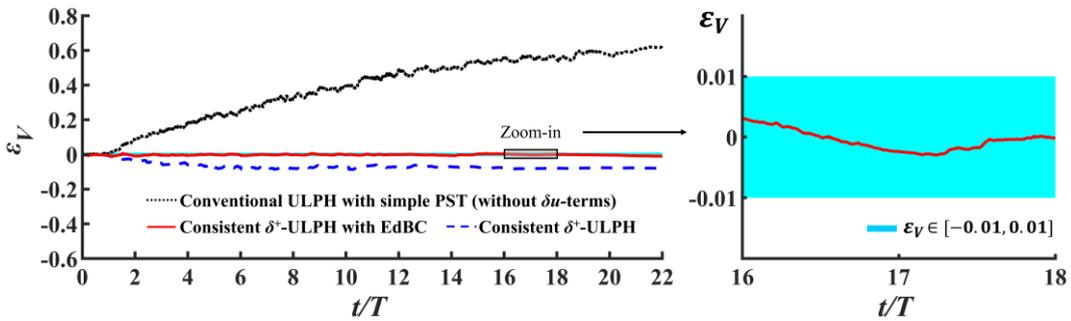

Fig. 24. Violent vertical sloshing: Volume errors of the ULPH models ($h$ = 2$\Delta x$).

As shown in Fig. 24, the consistent $\delta^+$-ULPH model with EdBC correction can obtain superior volume conservation, and the volume error $\varepsilon_V$ is lower than 0.01. Compared to the conventional ULPH model with simple PST, the consistent $\delta^+$-ULPH model significantly limits the volume expansion, attributed to $\delta\boldsymbol{u}$-terms in Eq. (29).

In addition, the energy dissipation is analyzed by a variable $\varepsilon_E$ through the following formula:

$$\varepsilon_E = \frac{E_t - E_0}{E_0 - E_\infty}, \qquad E_t = E_t^{\text{tank}} + E_t^{\text{fluid}} + E_t^{\text{spring}} \tag{48}$$

in which $E_t$, $E_0$, and $E_\infty$ represent the time evolution of total energy, initial total energy, and energy in steady state, respectively. $E_t^{\text{tank}}$ and $E_t^{\text{fluid}}$ are the mechanical energies of the tank and fluid, respectively. $E_t^{\text{spring}}$ is the potential energy of the spring.

Figure 25 compares the $\varepsilon_E$ obtained by the conventional ULPH with simple PST, consistent $\delta^+$-ULPH, and consistent $\delta^+$-SPH models. The time history of $\varepsilon_E$ of the consistent $\delta^+$-ULPH model agrees well with that of the consistent $\delta^+$-SPH model. According to Krimi et al. [71], non-physical volume expansion causes a non-physical energy increase. As shown in Fig. 24, the conventional ULPH model without $\delta\boldsymbol{u}$-terms exhibits visible volume expansion. As a result, due to the non-physical energy increase caused by volume expansion, the decay rate of $\varepsilon_E$ is visibly lower than that of the other two models.

In summary, benefitting from the $\delta\boldsymbol{u}$-terms in Eq. (29) and EdBC correction in Eq. (47), the consistent $\delta^+$-ULPH model achieves good conservation of volume and energy.

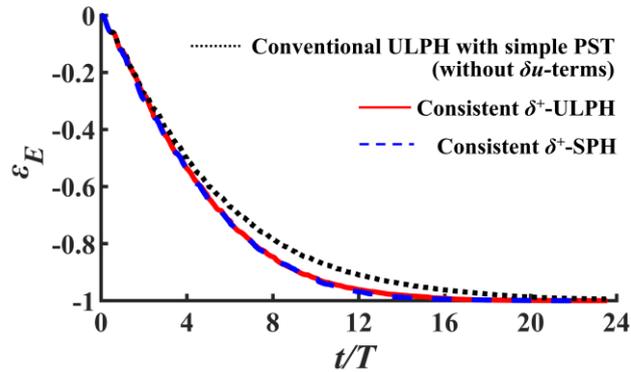

Fig. 25. Violent vertical sloshing: time history of the energy dissipation ($h = 2\Delta x$).

## 5. Conclusions

The conventional SPH model is afflicted by excessive numerical dissipation when the neighboring particles are insufficient. To address this, a novel consistent $\delta^+$-ULPH model is proposed in this paper, along with an extended support domain technique (ESDT), optimal matrix $\boldsymbol{D}$ for the velocity divergence, consistent particle shifting technique (PST), tensile instability control (TIC), and acoustic damper term. Model validation was carried out through five benchmarks, including rotation of a square fluid, oscillating droplet, attenuation of a standing wave, long-distance propagation of a regular wave, and violent vertical sloshing. The benchmark results of the conventional ULPH model [38] and the consistent $\delta^+$-SPH model [45] were computed for comparison to show the superiority of the consistent $\delta^+$-ULPH model.

Results indicate that the proposed consistent $\delta^+$-ULPH model can accurately simulate both gentle waves and violent sloshing flows and shows higher accuracy and lower numerical dissipation when using fewer neighboring particles, even in long-term and long-distance wave propagation simulations. The rotation of a square fluid calculated by the proposed model maintains good stability, even with large free-surface deformation, benefiting from the consistent PST. The oscillating droplet, standing wave, and long-distance wave propagation calculated by the proposed model show superior accuracy and energy conservation, attributed to the lower discrete error from ESDT. Additionally, the free-surface instability in long-term ULPH simulations is addressed by superior conservation from ESDT and the optimal matrix $\boldsymbol{D}$ for the velocity divergence. The violent vertical sloshing results of the proposed model agree well with the experiments, where the TIC and consistent PST contribute significantly to solving the tensile instability caused by the negative pressure field. In each benchmark, the pressure field of the present proposed model shows superior stability, benefiting from the acoustic damper term. In addition, with similar accuracy to the consistent $\delta^+$-SPH model, the computational efficiency of the consistent $\delta^+$-ULPH model is enhanced visibly because of fewer neighboring particles.

The limitations of this study are as follows. Firstly, this study conducts only 2D simulations. To further verify the present proposed model, 3D simulations are needed. Secondly, the present codes of the proposed model only run on the CPU. In large-scale and long-term 3D simulations, GPU parallel computing acceleration is necessary. Thirdly, the current consistent $\delta^+$-ULPH model is a single-phase flow model. In the future, the multi-phase $\delta^+$-ULPH scheme will be further studied.

**Acknowledgments**

This work was partially funded by the National Natural Science Foundation of China (Grant No. 52171329), the Guangdong Basic and Applied Basic Research Foundation (Grant No. 2024B1515020107), and the Characteristic Innovation Project of Universities in Guangdong Province (Grant No. 2023KTSCX005). The research leading to these results was also partially funded by the HASTA project (Grant No. 101138003) as part of the European Union Horizon research programme. Views and opinions expressed are however those of the authors only and do not necessarily reflect those of the European Union. Neither the European Union nor the granting authority can be held responsible for them.

**Appendix A. Derivation of the ULPH pressure gradient**

The discrete pressure gradient of the ULPH model is derived from the Peridynamics (PD) theory. The momentum equation of PD is written as follows:

$$\begin{cases} \rho_i \dfrac{D\boldsymbol{u}_i}{Dt} = \int_{H_i} \left( \boldsymbol{T}_i \langle \boldsymbol{R}_j - \boldsymbol{R}_i \rangle - \boldsymbol{T}_j \langle \boldsymbol{R}_i - \boldsymbol{R}_j \rangle \right) dV_j + \boldsymbol{f}_i \\ \boldsymbol{T}_i \langle \boldsymbol{R}_{ji} \rangle = W(\boldsymbol{R}_{ji}) \boldsymbol{P}_i \left( \boldsymbol{K}(\boldsymbol{R}_i) \right)^{-1} \boldsymbol{R}_{ji} \\ \boldsymbol{K}(\boldsymbol{R}_i) = \int_{H_i} W(\boldsymbol{R}_{ji}) \boldsymbol{R}_{ji} \otimes \boldsymbol{R}_{ji} dV_j \end{cases} \quad (A.1)$$

As a total Lagrangian formulation, Eq. (A.1) describes material motion relative to the initial configuration, using referential coordinates $\boldsymbol{R}$. $\boldsymbol{T}_i \langle \boldsymbol{R}_{ji} \rangle$ is the force state and $\boldsymbol{R}_{ji} = \boldsymbol{R}_j - \boldsymbol{R}_i$. $W(\boldsymbol{R}_{ji})$ and $\boldsymbol{K}(\boldsymbol{R}_i)$ are the kernel function and shape tensor in the initial configuration, respectively. $\boldsymbol{P}_i$ is the first Piola-Kirchhoff stress tensor. $\boldsymbol{f}_i$ is the external force. In ULPH, fluid simulations employ the updated Lagrangian approach, with Eq. (A.1) reformulated in the current (updated reference) configuration [40]. The momentum equation of ULPH is written as:

$$\begin{cases} \rho_i \dfrac{D\boldsymbol{u}_i}{Dt} = \int\limits_{H_i} \left( \boldsymbol{T}_i \langle \boldsymbol{r}_j - \boldsymbol{r}_i \rangle - \boldsymbol{T}_j \langle \boldsymbol{r}_i - \boldsymbol{r}_j \rangle \right) dV_j + \boldsymbol{f}_i \\ \boldsymbol{T}_i \langle \boldsymbol{r}_{ji} \rangle = W(\boldsymbol{r}_{ji}) \boldsymbol{\sigma}_i \left( \boldsymbol{M}(\boldsymbol{r}_i) \right)^{-1} \boldsymbol{r}_{ji} \\ \boldsymbol{M}(\boldsymbol{r}_i) = \int\limits_{H_i} W(\boldsymbol{r}_{ji}) \boldsymbol{r}_{ji} \otimes \boldsymbol{r}_{ji} dV_j \end{cases} \quad (A.2)$$

in which $W(\boldsymbol{r}_{ji})$ and $\boldsymbol{M}(\boldsymbol{r}_i)$ are the kernel function and shape tensor in the current configuration (i.e., updated configuration), respectively. To simplify the notation, $W_{ij}$ is used to replace $W(\boldsymbol{r}_{ji})$, which is consistent with the kernel function in SPH. $\boldsymbol{M}_i$ is used to replace $\boldsymbol{M}(\boldsymbol{r}_i)$. $\boldsymbol{\sigma}_i$ is the Cauchy stress. Since $\boldsymbol{r}_{ji} = \boldsymbol{r}_j - \boldsymbol{r}_i = -\boldsymbol{r}_{ij}$, Eq. (A.2) is rewritten as:

$$\rho_i \frac{D\boldsymbol{u}_i}{Dt} = \int\limits_{H_i} W_{ij} \left( \boldsymbol{\sigma}_i \boldsymbol{M}_i^{-1} + \boldsymbol{\sigma}_j \boldsymbol{M}_j^{-1} \right) \boldsymbol{r}_{ji} dV_j + \boldsymbol{f}_i \quad (A.3)$$

Without considering the viscosity force, the discrete pressure gradient of the ULPH is obtained from Eq. (A.3), written as:

$$\begin{cases} \langle \nabla p \rangle_i = \sum\limits_j W_{ij} \left( p_i \boldsymbol{M}_i^{-1} + p_j \boldsymbol{M}_j^{-1} \right) \boldsymbol{r}_{ji} V_j \\ \boldsymbol{M}_i = \sum\limits_j W_{ij} \boldsymbol{r}_{ji} \otimes \boldsymbol{r}_{ji} V_j \\ \boldsymbol{M}_j = \sum\limits_k W_{jk} \boldsymbol{r}_{kj} \otimes \boldsymbol{r}_{kj} V_k \end{cases} \quad (A.4)$$

**Appendix B. The effect of numerical diffusive terms and $\delta u$-terms on conservation**

In the consistent $\delta^+$-ULPH model, the numerical diffusive terms include the density diffusive term $\Phi_i$ in Eq. (16), the artificial/physical viscous term $\boldsymbol{F}_i^v$ in Eq. (17), and the acoustic damper term $\boldsymbol{F}_i^{ad}$ in Eq. (31). The $\delta\boldsymbol{u}$-terms considered in this Appendix include $\langle \nabla \cdot (\rho \delta\boldsymbol{u}) \rangle_i$ and $\langle \nabla \cdot (\rho \boldsymbol{u} \otimes \delta\boldsymbol{u}) \rangle_i$ in Eq. (29). These terms are always constructed as an anti-symmetry form with a symmetry matrix $\left( \boldsymbol{M}_i^{-1} + \boldsymbol{M}_j^{-1} \right)/2$. Considering global (integral) formulations, the integrations of the aforementioned terms are equal to 0 due to the anti-symmetry property, written as:

$$\begin{cases} \sum_i \Phi_i = 0, & \sum_i F_i^v = 0, & \sum_i F_i^{ad} = 0, \\ \sum_i \langle \nabla \cdot (\rho \delta u) \rangle_i = 0, & \sum_i \langle \nabla \cdot (\rho u \otimes \delta u) \rangle_i = 0 \end{cases} \quad (A.5)$$

As a result, the numerical diffusive terms and $\delta u$-terms do not introduce errors in the global mass and momentum and therefore ensure the conservation of mass and momentum.